\documentclass[twocolumn,showpacs,  preprintnumbers, superscriptaddress, amssymb,pra]{revtex4-1}

\usepackage{amssymb}
\usepackage{amsmath}
\usepackage{graphicx}
\usepackage{dcolumn}
\usepackage{bm}
\usepackage{verbatim}
\usepackage{array}
\usepackage{epsfig}
\usepackage{amsbsy}	
\usepackage{color}
\usepackage{soul}
\usepackage{url}
\usepackage{hyperref}
\usepackage{natbib}
\usepackage{longtable}

\usepackage[mediumqspace,mediumspace,amssymb,Gray]{SIunits}

\newcommand{\ds}{\displaystyle}
\newcommand{\ddsum}[1]{{\displaystyle \sum_{ #1 }}}

\newif\ifprintchanges

\printchangesfalse

\newcommand{\Blue}[1]{\ifprintchanges{\color{blue}{#1}}\else#1\fi}

\definecolor{darkgreen}{rgb}{0.0, 0.6, 0.6}

\def\bra#1{\mathinner{\langle{#1}|}}
\def\ket#1{\mathinner{|{#1}\rangle}}

\newcommand{\remred}{\ifprintchanges{\textcolor{red}{[...]}}\else{}\fi}

\begin{document}


\preprint{\hfill\parbox[b]{0.3\hsize}{ }}

\title{ Quantum interference effects  in laser spectroscopy of \\muonic hydrogen, deuterium, and helium-3}
 

%
        %

\author{Pedro Amaro}
\email{pdamaro@fct.unl.pt}
\affiliation{Laborat\'orio de Instrumenta\c{c}\~ao, Engenharia Biom\'edica e F\'isica da Radia\c{c}\~ao
(LIBPhys-UNL),~Departamento de F\'isica, Faculdade~de~Ci\^{e}ncias~e~Tecnologia,~FCT,~Universidade Nova de Lisboa,~P-2829-516 Caparica, Portugal.}

\author{Beatrice Franke}
\email{beatrice.franke@mpq.mpg.de}
\affiliation{Max-Planck-Institute of Quantum Optics, D-85748 Garching, Germany.}

\author{Julian J.\ Krauth}
\affiliation{Max-Planck-Institute of Quantum Optics, D-85748 Garching, Germany.}

\author{Marc Diepold}
\affiliation{Max-Planck-Institute of Quantum Optics, D-85748 Garching, Germany.}

\author{Filippo Fratini}
\affiliation{Atominstitut, Vienna University of Technology, A-1020 Vienna, Austria}

\author{Laleh Safari}
\affiliation{IST Austria, Am Campus 1, A-3400 Klosterneuburg, Austria.}

\author{\linebreak Jorge Machado}
\affiliation{Laborat\'orio de Instrumenta\c{c}\~ao, Engenharia Biom\'edica e F\'isica da Radia\c{c}\~ao
(LIBPhys-UNL),~Departamento de F\'isica, Faculdade~de~Ci\^{e}ncias~e~Tecnologia,~FCT,~Universidade Nova de Lisboa,~P-2829-516 Caparica, Portugal.}
\affiliation{Laboratoire Kastler Brossel, \'Ecole Normale Sup\'erieure, CNRS,  
Universit\'e P. et M. Curie -- Paris 6, Case 74; 4, Place Jussieu, F-75252 Paris CEDEX 05, France.}

\author{Aldo Antognini}
\affiliation{Institute for Particle Physics, ETH Zurich, CH-8093 Zurich, Switzerland.}
\affiliation{Paul Scherrer Institute, 5232 Villigen-PSI, Switzerland.}

\author{Franz Kottmann}
\affiliation{Institute for Particle Physics, ETH Zurich, CH-8093 Zurich, Switzerland.}

\author{Paul Indelicato}
\affiliation{Laboratoire Kastler Brossel, \'Ecole Normale Sup\'erieure, CNRS,  
Universit\'e P. et M. Curie -- Paris 6, Case 74; 4, Place Jussieu, F-75252 Paris CEDEX 05, France.}

\author{Randolf Pohl}
\affiliation{Max-Planck-Institute of Quantum Optics, D-85748 Garching, Germany.}

\author{Jos\'e Paulo Santos}
\affiliation{Laborat\'orio de Instrumenta\c{c}\~ao, Engenharia Biom\'edica e F\'isica da Radia\c{c}\~ao
(LIBPhys-UNL),~Departamento de F\'isica, Faculdade~de~Ci\^{e}ncias~e~Tecnologia,~FCT,~Universidade Nova de Lisboa,~P-2829-516 Caparica, Portugal.}

\date{Received: \today  }

\begin{abstract}
Quantum interference between energetically close states is theoretically investigated, with the state structure being observed via laser spectroscopy.
In this work, we focus on hyperfine states of selected hydrogenic muonic isotopes, and on how quantum interference affects the measured Lamb shift. 
The process of photon excitation and subsequent photon decay is implemented within the framework of nonrelativistic second-order perturbation theory. 
Due to its experimental interest, \remred calculations are performed for muonic hydrogen, deuterium, and helium-3. We restrict our analysis to the case of photon scattering by incident linear polarized photons and the polarization of the scattered photons not being observed\remred.
~We conclude that while quantum interference effects can be safely neglected in muonic hydrogen and helium-3, in the case of muonic deuterium there are resonances with close proximity, where quantum interference effects can induce shifts up to a few percent of the linewidth, assuming a pointlike detector.~However, by taking into account the geometry of the setup used by the CREMA collaboration, this effect is reduced to less \Blue{than} 0.2\% of the linewidth in all possible cases, which makes it irrelevant at the present level of accuracy. 

\end{abstract}


\maketitle


\section{Introduction}
\label{intro}

 \begin{figure*}[t]
  \centering
\includegraphics[clip=true,width=1.0\textwidth]{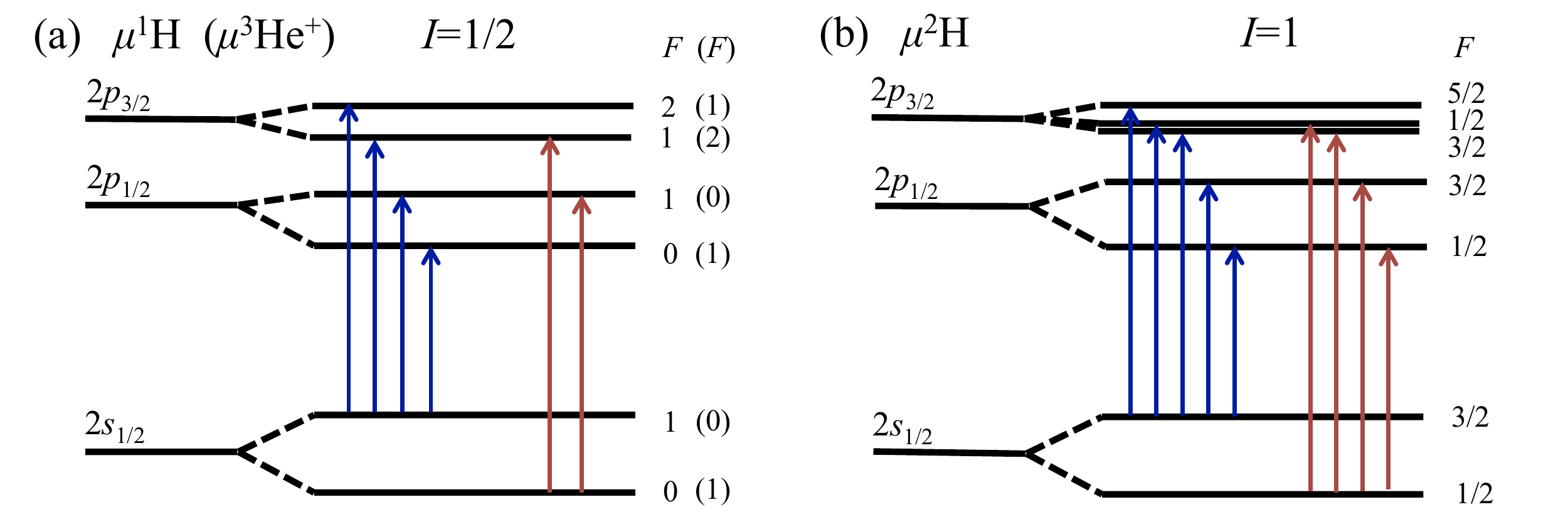}
\caption{ (Color online) $2s$ and $2p$ \Blue{level} structure \Blue{(not  in scale)} for $\mu ^1$H, $\mu ^3$He$^+$ (a)  and $\mu ^2$H (b). \remred 
 Allowed electric dipole \Blue{transitions} are also shown. \remred\Blue{$I$ and $F$ are the nuclear spin and total angular momentum, respectively.}
} 
\label{fig:transit}
\end{figure*}

Quantum interference (QI) corrections were introduced  in the seminal work of Low \cite{low1952}, where the quantum electrodynamics (QED) theory of the natural line profile in atomic physics was formulated. These corrections  go beyond the resonant approximation and set a limit for which a standard Lorentzian line profile can be used to describe a resonance. Often referred to as nonresonant corrections \cite{lkg1994, lsp2001,lga2009}, they contain the full quantum interference between the main resonant channel and other non-resonant channels, which leads to an asymmetry of the line profile. \Blue{Therefore, the fitting of spectroscopy data with Lorentzian} profiles becomes ambiguous, since it leads to energy shifts that depend on the measurement process itself \cite{lkg1994, lsp2001,lga2009, bup2013}. \Blue{A} careful analysis of the limits of the resonance approximation is \Blue{thus} mandatory for high-precision optical and microwave \Blue{spectroscopy} experiments. \remred 

The first calculation of QI was made for the Lamb shift in hydrogen-like (H-like) ions and for the photon scattering case \cite{low1952}. It was found to be relatively small, of the order of $\delta_{\tiny \mbox{QI}}/\Gamma \approx \alpha (\alpha Z)^2$, \Blue{compared to the linewidth $\Gamma$}. Here $\delta_{\tiny  \mbox{QI}}$ is the \Blue{line shift} due to QI, \remred $\alpha$ is the fine structure constant and $Z$ is the atomic number. Thus, for some time, little interest has been addressed to these corrections in optical measurements of H-like ions. Since the late 1990s, high-precision measurements of the $1s-2s$ transition frequency in hydrogen renewed the interest in these corrections \cite{hgw1999, nhr2000, pma2011}.  
Numerous theoretical calculations  of QI were then made for H-like ions  \cite{lsp2001, jem2002,lss2007, lga2009} with astrophysical interest \cite{kiv2008} and application to laser-dressed atoms \cite{jue2003}. QI has also been studied in other atomic systems and processes during the last decades, mainly because near and crossed resonances of hyperfine states can enhance the QI effects \cite{fra1961, wmp1982, scs2011}.  
\Blue{QI effects \cite{hoh2010, mhh2012, mhh2012b, mhh2014} have been shown to be responsible for discrepant measurements of the helium fine structure \cite{mhh2015, mhhb2015} and the lithium charge radii determined by the isotope shift \cite{bup2013}. The lithium
experiment \cite{bup2013} gives a beautiful experimental demonstration of the geometry
dependence of the QI effect. Very recently, QI effects in two-photon frequency-comb spectroscopy have been investigated, too \cite{mpb2014}.}

%
\begin{table}[t]
\caption{\label{tab:parame_hyper_2} Calculated energy differences of $\mu ^1$H, $\mu ^2$H and $\mu ^3$He$^+$ \cite{bor2012, akb2013} states relative to the lowest $n=2$ hyperfine state, which are $2s_{1/2} ^{F=0}$, $2s_{1/2} ^{F=1/2}$ and $2s_{1/2} ^{F=1}$, respectively. \Blue{ $\Gamma_{2p}$ is the respective approximate linewidth.}
} 
\begin{ruledtabular}
\begin{tabular}{llrr}
                     &   & GHz  & meV \\
		 \\[-2.0ex]  \cline{2-4} \\[-2.0ex]	\\[-2.0ex]	
$\mu ^1$H   & $2s_{1/2}(F=1)$&  5520   & 22.83  \\
	           & $2p_{1/2} (F=0)$&  51631  &  213.53\\
    	           & $2p_{1/2} (F=1)$&  53521  & 221.35\\
	           & $2p_{3/2} (F=1)$&  54616  & 225.88\\	
	           & $2p_{3/2} (F=2)$&  55402  & 229.12\\	
	           &    \Blue{$\Gamma_{2p}$~\cite{med1998} }&   \Blue{18.5}      &     \Blue{0.0765}\\ 
	            \\[-2.0ex]  \cline{2-4} \\[-2.0ex]	\\[-2.0ex]	
$\mu ^2$H & $2s_{1/2} (F=3/2)$&  1485 & 6.143 \\
	           & $2p_{1/2} (F=1/2)$&  49680 & 205.46\\
	           & $2p_{1/2} (F=3/2)$& 50182 & 207.54\\
	           & $2p_{3/2} (F=3/2)$& 52016 & 215.12\\	
	           & $2p_{3/2} (F=1/2)$& 52104 & 215.49\\	
	           & $2p_{3/2} (F=5/2)$& 52286 & 216.24 \\
	           &    \Blue{$\Gamma_{2p}$~\cite{med1998} }&   \Blue{19.5}      &     \Blue{0.0806}\\   
	            \\[-2.0ex]  \cline{2-4} \\[-2.0ex]	\\[-2.0ex]	
$\mu ^3$He & $2s_{1/2} (F=0)$&  41443  & 171.40\\
	           & $2p_{1/2}    (F=1)$&  311456 & 1288.08 \\
	           & $2p_{1/2} (F=0)$&  325633 & 1346.71\\
	           & $2p_{3/2} (F=2)$&  347860 & 1438.63\\	
	           & $2p_{3/2} (F=1)$&  353731 & 1462.92\\
	           &    \Blue{$\Gamma_{2p}$\footnote{\Blue{Calculated in the present work.}} }&   \Blue{318.7}      &     \Blue{0.1318}\\ 
\end{tabular}
\end{ruledtabular}
\end{table}

In this work, we calculate the QI \Blue{shifts}  for \Blue{$2s$$\rightarrow$$2p$} transitions in H-like muonic atoms \remred  
 with hyperfine structure.
The physical process considered here is the photon scattering of initial $2s$ states to final $1s$ states, with an incident photon energy that is resonant with intermediate \remred $2p$ states. \Blue{Figure~\ref{fig:transit} recalls the $2s\rightarrow2p$ level structures that are considered in this work. The level structures and linewidths \cite{bor2012,med1998} of the $2p$ states are given in Table~\ref{tab:parame_hyper_2}.} 
The theoretical formalism used here can be traced back to recent works \cite{saf2012b, sas2015,bup2013}. \remred

Hyperfine states in $\mu ^1$H are separated by \remred \Blue{several} hundred GHz \remred\cite{mar2005, mar2008b, bor2005,bor2012}, and have linewidths  of  a few tens of GHz \cite{med1998}. Thus, it is expected that QI plays a small role and cannot be responsible for the  \Blue{so-called {\it ``proton radius puzzle''}, where a discrepancy of four linewidths was observed in the experiments of the  Charge Radius Experiment with Muonic Atoms (CREMA) collaboration \cite{pan2010, ans2013}}. Nevertheless, these systematics need to be carefully evaluated and quantified, since they have contributions  similar to  small QED corrections \Blue{(e.g. sixth-order contributions \cite{bor2012}) and thus may impact  precise determinations of the proton charge radius}. \remred 
%
\Blue{Since} resonances in  $\mu ^3$He$^+$ \cite{abc2011, naa2012} have been measured recently, values of the corresponding QI contributions are presented here, \Blue{too}. In the case of $\mu ^2$H, there is a close energy proximity between the states $2p_{3/2}^{F=1/2}$ and $2p_{3/2}^{F=3/2}$ of  $\approx87$~GHz and\Blue{ hence,} it is expected that QI effects could be much higher.  \remred 
\section{Theory}
\label{sec:theor}
 
Photon scattering is a two-step process consisting of photon excitation with subsequent photon decay, which is formally equivalent to Raman anti-Stokes scattering. It is described by second-order theories \Blue{(e.g. Kramers-Heisenberg formula \cite{rlo2000}, or  S-matrix  \cite{alb1965}), which overall converge to the following scattering amplitude (velocity gauge and atomic units)} \Blue{from initial to final states} \cite{alb1965, rlo2000, saf2012b},
\Blue{\begin{eqnarray}
 \ds \mathcal{M}_{i \rightarrow f}^{ {\bm \varepsilon_1} {\bm\varepsilon_2} } =  
 \ds \ddsum{\nu}
\left[
\frac{ \bra{f} \alpha {\bm p }\cdot {\bm \varepsilon_2}
\ket{\nu}\bra{\nu} \alpha {\bm p }\cdot {\bm \varepsilon_1}  \ket{i}}{\omega_{\nu i}
-\omega_1 -i \Gamma_{\nu} /2 }  \right. \nonumber \\
 \left. + \frac{ \bra{f} \alpha {\bm p }\cdot {\bm \varepsilon_1}
\ket{\nu}\bra{\nu} \alpha {\bm p }\cdot {\bm \varepsilon_2}
 \ket{i}}{\omega_{\nu i}+\omega_2 -i \Gamma_{\nu} /2 }
 \right], \hspace{-0.25cm}  
\label{Mfi}
\end{eqnarray}}\noindent
where $\ket{i}$, $\ket{\nu}$, and $\ket{f}$ represent the initial, intermediate and final hyperfine states of  the muonic atom or ion. 
$\omega_{\nu i} = E_\nu - E_i $ is the transition frequency between  $\ket{\nu}$ and $\ket{i}$. \remred\Blue{The dipole approximation, $\alpha {\bm p }\cdot {\bm \varepsilon_\gamma}$  ($\gamma=1,2$) is used, where, ${\bm p}$ is the linear momentum operator and ${\bm \varepsilon_\gamma}$ is the incident (scattered) photon polarization.}
%
%
The summation over the intermediate states \Blue{$\ket{\nu}$} runs over all solutions of the Dirac spectrum of the muonic atom (ion) with hyperfine structure. 
 \Blue{All states are considered with a well-defined total atomic angular momentum $F$, projection along the quantified axis $m$, and total orbiting particle angular momentum $J$ \cite{sas2014}; thus the contribution of off-diagonal terms (\Blue{mixing between the $2p_{1/2}^{F=1}$ and $2p_{3/2}^{F=1}$} states) \cite{pac1996} is considered null. $\Gamma_{\nu}$ is the full width at half maximum
for an isolated Lorentzian line of the excited state, where we assume $\Gamma_{2p_j^{F}}\equiv\Gamma_{2p}$.} \remred

The \remred incident photon energies studied in this work comprise the near-resonant region of the  \Blue{$2s\rightarrow2p$ transitions}. This includes \Blue{only the} resonances illustrated in Fig.~\ref{fig:transit}.  \remred\Blue{Hence, we restrict the summation $\nu$ to the $2p$  states of the terms $\omega_{2s2p} -\omega_1$ [first part of the right side of Eq.~\eqref{Mfi}].} \remred 

 \begin{figure}[t]
  \centering
\includegraphics[clip=true,width=1.0\columnwidth]{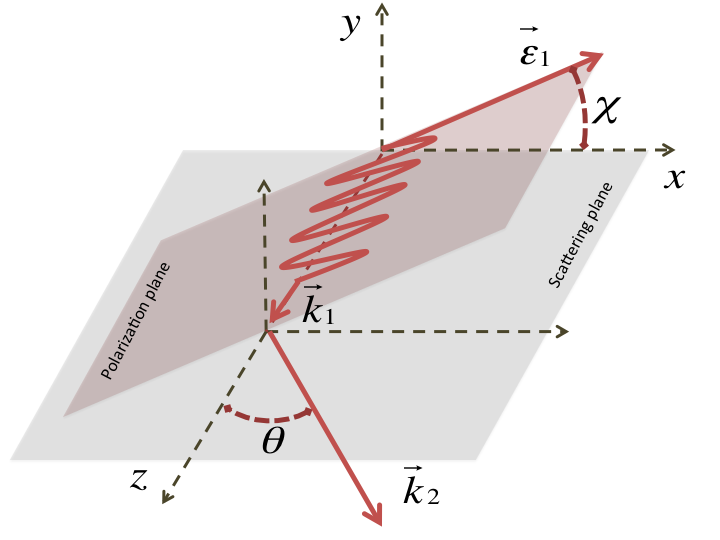}
\caption{ (Color online) Adopted geometry for photon scattering of incident linear polarized photons with \Blue{momentum} ${\bm k_1}$ and polarization ${\bm \varepsilon_1}$ \Blue{and scattered photon momentum ${\bm k_2}$, uniquely defined by $\theta$}. 
The \Blue{scattered} photon's polarization is not observed \Blue{in our measurements and is thus} not illustrated. \remred
} 
\label{fig:geom}
\end{figure}

Energy conservation leads to \Blue{$E_i- E_f =\omega_2 - \omega_1$} \cite{saf2012b} between \Blue{the initial ($E_i$) and final ($E_f$)}  energy states and the energy of the incident \Blue{($\omega_1$)} and scattered \Blue{($\omega_2$)} photons;
thus only one of the photon
energies is independent. Using this relation, it is convenient to \Blue{introduce} the 
energy sharing parameter $u = \omega_1/\omega_r$ \Blue{defined by the fraction of the incident photon energy relative to the lowest resonant energy $\omega_r$ \remred of a given muonic atom with initial $F_i$ (see Fig.~\ref{fig:transit}).}

\Blue{Motived by the experimental configuration, we consider incident photons having linear polarization} and
non-observation of \Blue{scattered} photon's polarization, \Blue{as} illustrated in Fig.~\ref{fig:geom} \remred \Blue{\footnote{Elliptical polarization of the incident photon can only decrease the QI shift and will be reported elsewhere.}}. \remred If we define the scattering plane, containing both \Blue{photon momenta} (${\bm k_1}$ and ${\bm k_2}$), \remred then a single polar angle $\theta$ is sufficient for describing the angular distribution \Blue{of ${\bm k_2}$.}   \remred
%
%

%
 \begin{figure*}[t]
  \centering
\includegraphics[clip=true,width=1.0\textwidth]{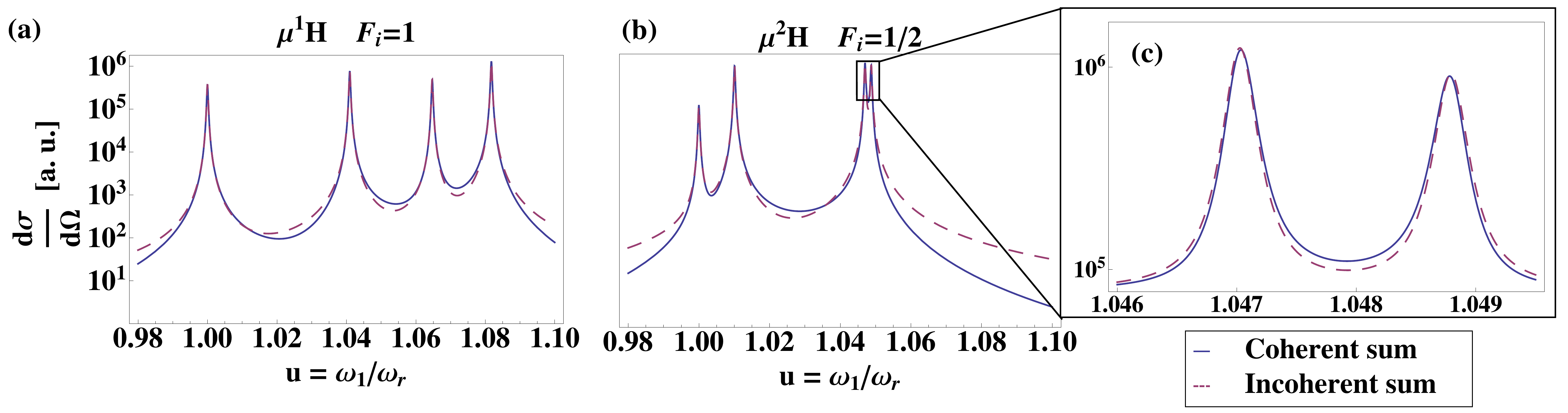}
\caption{ (Color online) Scattering \Blue{differential} cross-section \remred \eqref{eq:Msimpli_2} at $\theta=90$\degree and $\chi=0$\degree~ for  $\mu ^1$H \Blue{(a)} \remred  and $\mu ^2$H \Blue{(b)} \remred \Blue{versus photon frequency}. A zoom plot of the two close-lying resonances for $\mu ^2$H is given in the right panel (c). 
 Solid blue lines correspond to  the full evaluation of Eq.~\eqref{eq:Msimpli_2}, while the dashed red curves \Blue{account for only the first term in the right side of Eq.~\eqref{eq:Msimpli_2} (sum of Lorentzian profiles)}. \remred 
%
%
%
} 
\label{fig:plots_scater}
\end{figure*}

The corresponding differential cross section of the amplitude \Blue{in Eq.~\eqref{Mfi} for} all the mentioned \remred approximations is given by \cite{alb1965}

\Blue{
\begin{eqnarray}
\frac{d\sigma}{d\Omega}(u, \theta, \chi) =  
 \frac{1}{(2F_{i}+1)}\sum_{\tiny \begin{array}{c} m_{i},F_f,m_{f}, J_f \\ \bm{\varepsilon_{2}} \end{array} }\left|\mathcal{M}_{i\rightarrow f}^{\bm{\varepsilon_{1}},\bm{\varepsilon_{2}}}\right|^{2}   \approx    \hspace{1.0 cm} \nonumber \\
 \frac{ \omega_r^2 u u_{fi}^3 }{(2F_{i}+1)}
 \sum_{\tiny \begin{array}{c} m_{i},F_f,m_{f}, J_f \\ \bm{\varepsilon_{2}} \end{array} }
 \left| \sum_{F_\nu, m_\nu, J_\nu} \frac{D_{F_i m_i J_i}^{F_\nu m_\nu J_\nu}  \left(D_{F_\nu m_\nu J_\nu}^{F_f m_f J_f } \right)^*}{\bar{\omega}_{\nu i} - u -i \bar{\Gamma}_{\nu}/2} \right|^{2}
 ,\hspace{0.305 cm}\nonumber \\
 \label{eq:Msimpli}
\end{eqnarray} }\noindent
where \remred \Blue{$\bar{\omega}_{\nu i}=\omega_{\nu i}/ \omega_r$, $\bar{\Gamma}_\nu=\Gamma_\nu/ \omega_r$ and $u_{fi}=u+\bar{\omega}_{fi}$}. 
In Eq.~\eqref{eq:Msimpli}, it is assumed that the initial state of the atom is unpolarized and that the level and magnetic sublevels of
the final state, as well as the scattered photon's polarization \Blue{(${\bm \varepsilon_2}$)} remains unobserved in the scattering process. 
%
%
%
%
\Blue{$D_{F' m' J'}^{F m J} $ are the dipole matrix elements (length gauge). }
%
%
%
  Equation~\eqref{eq:Msimpli} can be further rearranged  \Blue{as} a sum of Lorentzian components $\Lambda_{J_{i}J_{\nu}}^{F_{i} F_{\nu}}( \theta, \chi)$, and cross terms $\Xi_{J_{i}J_{\nu} J_{\nu}'}^{F_{i} F_{\nu}  F_{\nu}'}( \theta, \chi)$, \Blue{similar to} Ref.~\cite{bup2013}. \Blue{For our particular geometry and atomic system,} the result is given by 
\begin{widetext}
\begin{equation}
\frac{d\sigma}{d\Omega}(u, \theta, \chi) = 
\frac{ \omega_r^2 u u_{fi}^3 \mathcal{S}_{f \nu i}^2 }{(2F_{i}+1)} 
 \left(
 \sum_{\small F_\nu,  J_\nu}  \frac{  \Lambda_{J_{i}J_{\nu}}^{F_{i} F_{\nu}}( \theta, \chi) }{(\bar{\omega}_{\nu i} - u)^2 + (\bar{\Gamma}_{\nu}/2 )^2 } +   \sum_{(F'_\nu,  J'_\nu) > (F_\nu,  J_\nu)}   \frac{\Xi_{J_{i}J_{\nu} J_{\nu}'}^{F_{i} F_{\nu} F_{\nu}'}( \theta, \chi)}{    (\bar{\omega}_{\nu i} - u - i \bar{\Gamma}_{\nu}/2) (\bar{\omega}_{\nu'  i} - u + i \bar{\Gamma}_{\nu}/2)}
 \right) ~,
 \label{eq:Msimpli_2}
\end{equation}
\end{widetext}
where the second summation \Blue{over} $F'_\nu$ and $ J'_\nu$ \Blue{runs} for non-repeated values of $F_\nu$ and $ J_\nu$ of the first summation. The quantities defined by  
\begin{eqnarray}
\Lambda_{J_{i}J_{\nu}}^{F_{i} F_{\nu}}( \theta, \chi)=   \sum_{m_{i},F_f,m_{f}, J_f, {\bm \varepsilon_{2} } }\left| \Omega_{J_{i}J_{\nu}J_{f}}^{F_{i} F_{\nu}F_{f}}( \theta, \chi, {\bm \varepsilon_{2} } ) \right|^2~, \hspace{1.0 cm}   \nonumber \\
\mbox{and} \hspace{8.0 cm}  \nonumber \\
\Xi_{J_{i}J_{\nu} J_{\nu}'}^{F_{i} F_{\nu}  F_{\nu}'}( \theta, \chi)=   \hspace{5.75 cm}   \nonumber \\  
2 \mbox{Re} \left[  \sum_{ \tiny \begin{array}{c} m_{i },F_f\\ m_{f} ,J_f, {\bm \varepsilon_{2} } \end{array}}  
\Omega_{J_{i}J_{\nu}J_{f}}^{F_{i}F_{\nu}F_{f}}( \theta, \chi, {\bm \varepsilon_{2} } )  
 \left(\Omega_{J_{i}J'_{\nu}J_{f}}^{F_{i} F'_{\nu}F_{f}}( \theta, \chi, {\bm \varepsilon_{2} } )\right)^*  \right]~,  \nonumber \\
%
%
\label{eq:Xi}
\end{eqnarray} 
contain all the polarization and geometrical dependencies.  \Blue{$D_{F' m' J'}^{F m J} $, $\Omega_{J_{i}J_{\nu}J_{f}}^{F_{i}F_{\nu}F_{f}}( \theta, \chi, {\bm \varepsilon_{2} } )$ and $\mathcal{S}_{f \nu i}$ are given in the Appendix}.

The differential cross section of Eqs.~\eqref{eq:Msimpli} and \eqref{eq:Msimpli_2} contains a {\it coherent} summation over resonant excitation channels; thus it takes into account channel-interference between neighboring resonances. However, as can be observed in Eq.~\eqref{eq:Msimpli_2}, if cross terms $\Xi$ were removed, it reduces to an {\it incoherent} sum of  independent Lorentzian \Blue{profiles.  The} QI effects are thus included in \Blue{those} cross terms. \remred

\section{Results and Discussion}
\label{sec:res_dis}


In this section, we present and discuss results for the QI contribution in several muonic atoms taking into account Eqs.~\eqref{eq:Msimpli_2} and \eqref{eq:Xi}, \Blue{first assuming a pointlike detector and later the CREMA geometry}. The influence of the geometric and polarization conditions on the QI is well described in Ref.~\cite{bup2013} and is reproduced in the present work.  We thus restrict our geometrical settings to the perpendicular observation ($\theta=90^o$) of scattered photons and to the case of horizontally and vertically polarized photons with respect to the scattering plane ($\chi=0^o$ and $\chi=90^o$).  
Figure \ref{fig:plots_scater} displays the scattering cross section for the $2s \rightarrow 2p \rightarrow1s$ \Blue{processes in} $\mu^1$H and $\mu^2$H, \Blue{on one hand,} having the full coherent summation \Blue{of Eq.~\eqref{eq:Msimpli} (i.e.~with QI), on the other hand,} having the summation restricted to only the Lorentzian terms [neglecting cross-terms \Blue{in Eq.~\eqref{eq:Msimpli_2}}]. The peaks correspond to the respective \Blue{transitions} shown in Fig.~\ref{fig:transit}. As it is observed, the influence of QI is more noticeable in regions between resonances, where no dominant excitation channels exist. Close to resonances, the influence of QI is \Blue{approximately} equivalent to shifting the peak position, as shown in the zoom plot of Fig.~\ref{fig:plots_scater}.

\Blue{We determine this shift in each resonance by generating a pseudo spectrum that  follows the theoretical profile of Eq.~\eqref{eq:Msimpli_2} and fitting it with an incoherently sum of Lorentzians (as performed in the data analysis of the CREMA experiments). Fits are done using the ROOT/MINUIT package \cite{min2015}. \remred
%
All fit parameters (position, amplitude, and linewidth) are free fit parameters for each transition.
The fit range is chosen sufficiently large, such that the fit results do not depend on it.}
%
%
The \remred shifts of the fitted resonance position, $\delta_{\tiny \mbox{QI}}$, \Blue{normalized to $\Gamma_{2p}$}, for each resonance and muonic atoms are given in Table~\ref{tab:shifts}. Overall, with the exception of some resonances in $\mu^2$H, QI produces relative shifts $\delta_{\tiny \mbox{QI}}$ less than 3\% \Blue{of $\Gamma_{2p}$}.
For $\mu^1$H and $\mu^3$He$^+$, the shifts are of the order of $\sim0.2\%-1.7$\% ($\sim$36~MHz$-310$~MHz) and $\sim0.2\%-3.2$\% ($\sim0.6$~GHz$-10$~GHz)  \Blue{of their linewidths}, respectively.
%
%
%
\begin{table}[t]
\caption{\label{tab:shifts} Shift of the line center due to QI contribution for $\mu ^1$H, $\mu ^3$He$^+$ and $\mu ^2$H. \Blue{$\ket{i}$} and \Blue{$\ket{\nu}$} stand for the initial and resonant atomic state, respectively. Values of $\delta$ \Blue{divided by $\Gamma_{2p}$} are given for a pointlike detector with  vertical ($\delta_{\tiny \mbox{QI}}^{\perp}$, ~\Blue{$\chi=90$\degree,~$\theta=90$\degree}) and horizontal ($\delta_{\tiny \mbox{QI}}^{\parallel}$, \Blue{$\chi=0$\degree,~$\theta=90$\degree}) polarization cases, as well as for the CREMA setup geometry ($\delta_{\tiny \mbox{QI}}^{*}$).
} 
\begin{ruledtabular}
\begin{tabular}{lllD{.}{.}{3}D{.}{.}{3}D{.}{.}{3}}
%
%
& \Blue{$\ket{i}$} & \Blue{$\ket{\nu}$}   &   \multicolumn{1}{l}{ $\Blue{\delta_{\tiny \mbox{QI}}^{\perp}}$ (\%) }  &   \multicolumn{1}{l}{ $\Blue{\delta_{\tiny \mbox{QI}}^{\parallel}}$ (\%) }&   \multicolumn{1}{l}{ $\Blue{\delta_{\tiny \mbox{QI}}^{*}}$ (\%) } \\
$\mu ^1$H & $2s_{1/2} ^{F=0}$ & $2p_{1/2} ^{F=1}$&    -0.8   & 1.6 & -0.02\\
	           &  &  $2p_{3/2} ^{F=1}$ &    0.4   &   -1.7 & 0.01\\
	           \\[-2.0ex]  \cline{2-6} \\[-2.0ex]	\\[-2.0ex]	
	           & $2s_{1/2} ^{F=1}$  &  $2p_{1/2} ^{F=0}$ &  -0.2  & 0.5& -0.01\\
	           &				&  $2p_{1/2} ^{F=1}$& -0.6 & 1.2& -0.01\\	
	     	&			          & $2p_{3/2} ^{F=1}$&  -0.5 & 0.7& -0.01\\    
		&			          & $2p_{3/2} ^{F=2}$ &   0.3 & -1.2& 0.01\\
	     \\[-2.0ex]        \hline	 \\[-2.0ex] 
$\mu ^3$He$^+$ &  $2s_{1/2} ^{F=0}$& $2p_{1/2} ^{F=1}$& -0.4 & 0.7& 0.00\\     
	         &			  & $2p_{3/2} ^{F=1}$&  0.2& -0.6 &   0.05\\
	           \\[-2.0ex]  \cline{2-6} \\[-2.0ex]	\\[-2.0ex]	
	           & $2s_{1/2} ^{F=1}$  & $2p_{1/2} ^{F=1}$& -0.3 & 0.5   & -0.01\\
	           &				& $2p_{1/2} ^{F=0}$ & -0.6   &  1.4    & -0.01\\	
	           &			          & $2p_{3/2} ^{F=2}$&  -0.1 & 0.4 & 0.00\\
	     	&			          & $2p_{3/2} ^{F=1}$&   2.2 & -3.2  & -0.02\\	
	 \\[-2.0ex] 	   \hline		 \\[-2.0ex] 	  
$\mu ^2$H & $2s_{1/2} ^{F=1/2}$ & $2p_{1/2} ^{F=1/2}$ &  -0.3 &  0.7 & -0.01\\
	         		 & & $2p_{1/2} ^{F=3/2}$&    -0.5  &  1.0 & -0.01\\
	            	 & & $2p_{3/2} ^{F=3/2}$&  -3.2 & 6.7.   &  -0.08\\	 
	            	 & & $2p_{3/2} ^{F=1/2}$&   4.4  & -8.0  & 0.11\\  
		  \\[-2.0ex]  \cline{2-6} \\[-2.0ex]	\\[-2.0ex]	
		 & $2s_{1/2} ^{F=3/2}$ & $2p_{1/2} ^{F=1/2}$&   -0.3 &  0.7 &-0.01\\
		 	 & & $2p_{1/2} ^{F=3/2}$&   -0.4 &   0.9   & -0.01\\
	            	 & & $2p_{3/2} ^{F=3/2}$&  -1.2 & 1.9  & -0.03\\	
	            	 & & $2p_{3/2} ^{F=1/2}$&  -4.9 & 12.3 &-0.13\\	
		          & & $2p_{3/2} ^{F=5/2}$&  0.9   & -2.7   & 0.03\\	
\end{tabular}
\end{ruledtabular}
\end{table}
 The observed discrepancy  (proton radius puzzle) of $\sim75$~GHz ($0.31$~meV) \cite{pan2010, ans2013} \Blue{at the $2p_{3/2}^{F=2}$ resonance} corresponds to 4  linewidths. \Blue{Hence, it is} much larger than any possible QI contribution. Moreover, \Blue{this} resonance is $\sim$7 times more intense than the closest resonance $2p_{3/2}^{F=1}$ (see Fig.~\ref{fig:plots_scater}), which minimizes the QI shift in this resonance. \remred
%
%
%
 The values presented in Table \ref{tab:shifts} for  $\mu^1$H and  $\mu^3$He$^+$ have the same order as the respective ones given by the rule-of-thumb for distant resonances ($\Blue{\delta_{\mbox{\tiny QI}}\sim \Gamma_{2p}}^2/4 \Delta$ with $\Delta$ being the energy difference between two resonances) \cite{hoh2010}. Apart from this, relatively low intensity resonances, like $2p_{3/2}^{F=1}$ in $\mu ^3$He$^+$, can have higher QI contributions due to a high intensity resonance nearby.

On the other hand, the resonances $2p_{3/2}^{F=1/2}$ and $2p_{3/2}^{F=3/2}$ in $\mu^2$H are more sensitive to QI effects not only due to their close proximity ($87$~GHz), but also due to the intensities being comparable within a factor of  $\sim$0.7. In this case, the QI shifts can be up to 12\% \Blue{of $\Gamma_{2p}$ ($\sim2$~GHz)}. \remred%

Applying the previously calculated \Blue{cross sections} to the geometry of the CREMA setup leads to considerable cancellations of the quantum interference effect. \remred

\begin{figure}[b]
\centering
\includegraphics[width=\columnwidth]{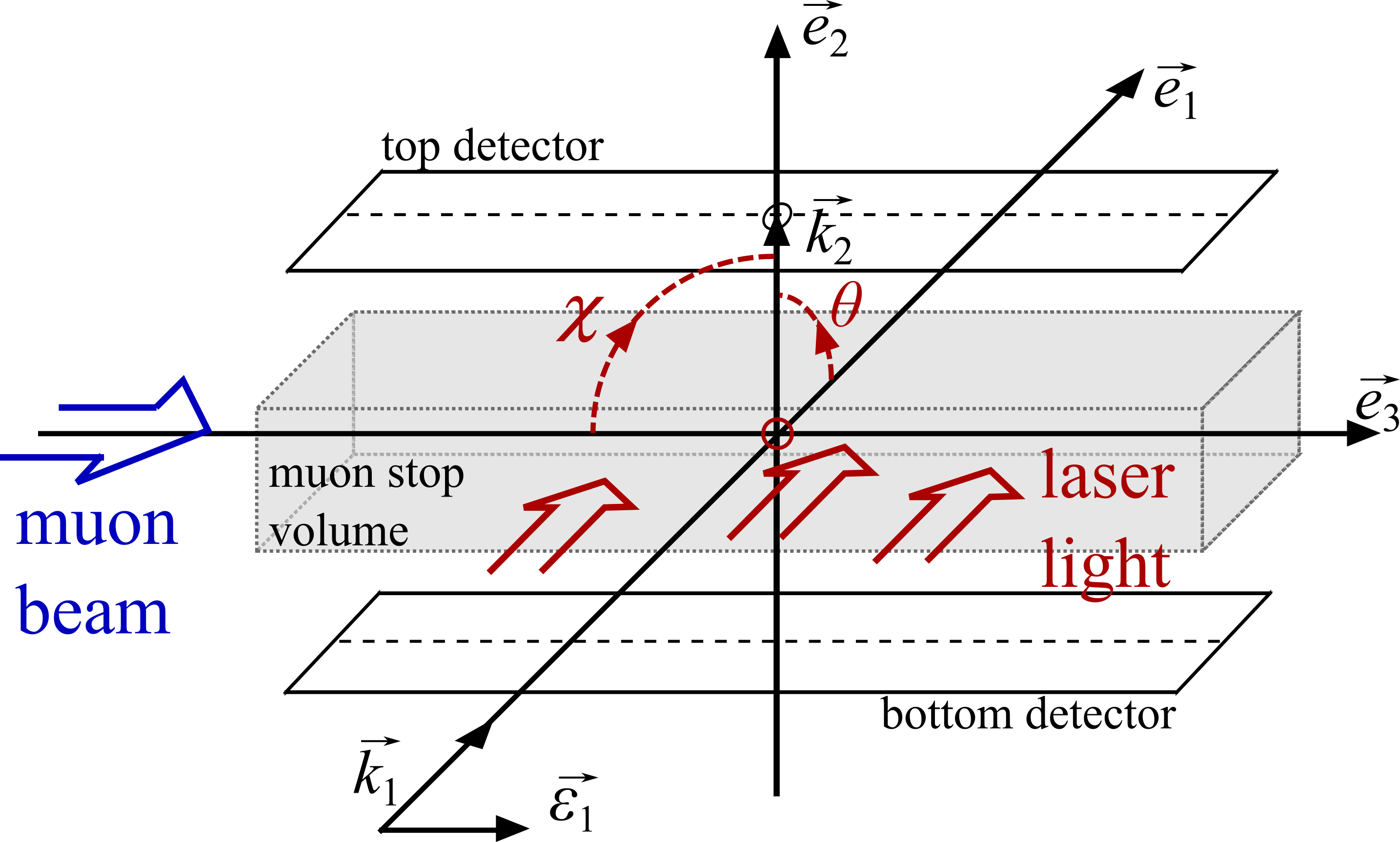}
\caption{
(Color online) Simplified view of the CREMA target volume.
Muons enter the gas target from the left (blue arrow). \remred
The gray cuboid indicates the muon stop volume which is illuminated by a laser pulse propagating inside a multipass cavity (red arrows). 
The incoming direction and polarization are indicated by $\boldsymbol{k_1}$ and $\boldsymbol{\varepsilon_1}$. \remred 
\Blue{The pointlike detector assumption treats the case where the muonic atom is at the center (red circle)
and the scattered photon $\boldsymbol{k_2}$ is emitted along $\boldsymbol{e_2}$ and hits the top detector in the center (black circle).}
}%
\label{fig:crema_geo}%
\end{figure}

\Blue{
Figure~\ref{fig:crema_geo} sketches the experimental geometry. Muonic atoms (ions) are formed in an
elongated gas volume of 5$\times12\times$190~mm$^3$ that is illuminated from the side using
a pulsed laser \cite{aab2005, asa2009} and a multipass cavity \cite{vda2014}. The $2p\rightarrow1s$ photons emitted
after laser-induced $2s\rightarrow2p$ transitions are detected by two
x-ray detectors (14$\times$150~mm$^2$ active area each) \cite{laa2005,faa2007,dfl2015} placed 8~mm above and below the muon beam axis.

 }

The simplest situation is that \Blue{the $2s\rightarrow2p$ excitation happens} in the center of the target (red circle in Fig.~\ref{fig:crema_geo}), and the resulting $\boldsymbol{k_2}$ \Blue{photon} is detected in the center of the top detector (black circle).
This corresponds to the \Blue{pointlike detector} case of $\theta=90^{\circ}$ and $\chi=90^{\circ}$ \Blue{($\delta_{\tiny \mbox{QI}}^{\perp}$ in Table~\ref{tab:shifts})}. \remred

\Blue{
However, the $2s\rightarrow2p$ excitation takes place anywhere in the muon stop volume, and
the photons of the $2p\rightarrow1s$ decay are detected anywhere on the detectors surfaces.
We consider here the laser's propagation (${\bm k_1}$) and polarization (${\bm \varepsilon_1}$)
directions being along ${\bm e_1}$ and ${\bm e_3}$, respectively.
Integrating Eq.~\eqref{eq:Msimpli_2} over all possible angles $\chi$ accepted by the detector, \remred while averaging over the muon stopping volume 
 results in a considerable reduction of the
observed QI effect. Again, we create pseudo data for the real geometry, fit the
resonances with a simple sum of Lorentzians, and determine the resulting shift
$\delta^*_{QI}$ of the line centers. 
We notice that taking into account the scattered photons at $\theta\ne90$\degree~and also the inhomogeneous muon stop probability, hardly affects the final results.
As can be seen in Table~\ref{tab:shifts}, these shifts are
much lower than the experimental accuracy of a few \% of the linewidth for all muonic atoms considered here \cite{pan2010, ans2013}.
}

\section{Conclusion}  
\label{sec:sum}


\remred \Blue{We quantified the line shift caused by quantum interference for} $\mu^1$H, $\mu^2$H and $\mu^3$He$^+$ resonances, \Blue{assuming first} a pointlike detector. For $\mu^1$H, the resulting \Blue{shifts} are small. Hence, quantum interference cannot be the source of the proton radius puzzle, which requires a shift of the resonance in $\mu^1$H by four linewidths \cite{pan2010, ans2013}. On the other hand, the influence of quantum interference \Blue{for some} resonances of $\mu^2$H can be as large as 12\% of the linewidth for a pointlike detector. \remred

\Blue{However, we verified} that even for those large QI \Blue{shifts,} 
obtained assuming a pointlike detector, the angular averaging caused by the large \Blue{acceptance angle of the photon} detector and \Blue{the size of} the muon stop volume  in the CREMA experiment  significantly reduces this effect to negligible values \Blue{at the present level of accuracy}.  

%
\begin{acknowledgments}

This research was supported in part by Funda\c{c}\~{a}o para a Ci\^{e}ncia e a Tecnologia (FCT), Portugal,
through the projects No. \emph{PEstOE/FIS/UI0303/2011} and \emph{PTDC/FIS/117606/2010}, financed by the European
Community Fund FEDER through the COMPETE.
P.~A. and J.~M. acknowledge the support of the FCT, under Contracts No. \emph{SFRH/BPD/92329/2013} and \emph{SFRH/BD/52332/2013}.  
B.\ F., J.\ J.\ K., M.\ D. and R.~P.\ acknowledge the support from the European Research Council (ERC) through StG.\ \#\emph{279765}. 
F.F. acknowledges support by the Austrian Science Fund (FWF) through the START grant \emph{Y 591-N16}. 
L. S. acknowledges
financial support from the People Programme (Marie Curie Actions) of the European Union's Seventh Framework Programme (\emph{FP7/2007-2013}) under REA Grant Agreement No. [\emph{291734}]. A.~A. acknowledge the support from  SNF \emph{200020\_159755}. \Blue{B.\ F. and R.\ P. would like to thank A. Beyer, L. Maisenbacher and Th. Udem for illuminating discussions.}

\end{acknowledgments}

%
\appendix

\begin{table*}[t]
\caption{\label{tab:coe_para_del_Xi} Values of the coefficients $a_0$, $a_2$ and $b_2$ corresponding to the parametrizations $\Lambda_{J_{i}J_{\nu}}^{F_{i} F_{\nu}}( \theta, \chi)=a_0 + a_2 P_2 (\cos \gamma)$ and $\Xi_{J_{i}J_{\nu}J_{\nu'}}^{F_{i} F_{\nu}F_{\nu'}}( \theta, \chi)=b_2 P_2 (\cos \gamma)$.
} 
\begin{ruledtabular}
\begin{tabular}{lllllllll}
$I$ & $F_i$ & $F_\nu$  & $J_\nu$ & $F_{\nu'}$ & $J_{\nu'}$  & $a_0$& $a_2$ & $b_2$\\
1/2 & 0 & 1 & 1/2 & 1 & 3/2 & 2/81 & 0 &  -4/81 \\   
 & 0 & 1 & 3/2 & - & - & 4/81 & -2/81 & - \\  
 & 1 & 1 & 1/2 & 1 & 3/2 & 4/81 & 0 &  -2/81 \\  
 & 1 & 1 & 1/2 & 2 & 3/2 & 4/81 & 0 &   -2/27 \\   
 & 1 & 1 & 3/2 & 2 & 3/2 & 2/81 & 1/162   & -1/27 \\   
 & 1 & 2 & 3/2 & 1 & 3/2 & 10/81 & -7/162 &  - \\     
 \\[-2.0ex]   \hline \\[-2.0ex]	
 
1 & 1/2 & 1/2 & 1/2 & 3/2 & 3/2 & 4/729 & 0 &  -8/729 \\  
 & 1/2 & 1/2 & 3/2 & 3/2 & 1/2 & 32/729 & 0 &  -32/729 \\  
 & 1/2 & 3/2 & 3/2 & - & - & 40/729 & -4/729 &  - \\  
 & 3/2 & 1/2 & 1/2 & 3/2 & 3/2 & 32/729 & 0 &  -32/3645 \\   
 & 3/2 & 1/2 & 1/2 & 5/2 & 3/2 & 32/729 & 0 &  -32/405 \\  
 & 3/2 & 1/2 & 3/2 & 3/2 & 1/2 & 4/729 & 0 &  -4/729 \\  
 & 3/2 & 1/2 & 3/2 & 3/2 & 3/2 & 4/729 & 0 &  16/3645 \\  
 & 3/2 & 1/2 & 3/2 & 5/2 & 3/2 & 4/729 & 0 &  -4/405 \\  
 & 3/2 & 3/2 & 1/2 & 3/2 & 3/2 & 40/729 & 0 &  -128/3645 \\   
 & 3/2 & 3/2 & 1/2 & 5/2 & 3/2 & 40/729 & 0 &  -28/405 \\  
 & 3/2 & 3/2 & 3/2 & 5/2 & 3/2 & 32/729 & 64/18225 &  -112/2025 \\  
 & 3/2 & 5/2 & 3/2 & - & - & 4/27 & -28/675 &  - \\  
 	         	         
\end{tabular}
\end{ruledtabular}
\end{table*}

\section{DIPOLE MATRIX ELEMENTS}

The dipole matrix elements $D_{F' m' J'}^{F m J }$ in Eq.~(2) are evaluated using standard angular reduction methods, which can start by simply expanding the product of the photon polarization and the position vector ($\bm{\varepsilon_{\gamma}} \cdot \bm{r} $) in a spherical basis, i.e.,
\begin{eqnarray}
D_{F' m' J'}^{F m J} &=&  \bra{\beta' F' m' J'} \bm{\varepsilon_{\gamma}} \cdot \bm{r}  \ket{ \beta F m J} = \nonumber \\
&&\sum_{\lambda=-1}^1 (-1)^\lambda  \varepsilon_\gamma^{-\lambda}  \bra{\beta' F' m' J'} r_\lambda  \ket{ \beta F m J}.\nonumber \\
\label{eq:matr_ele}
\end{eqnarray}
where $\beta$ contains all additional quantum numbers of the atomic state besides $F$, $m$ and $J$.
Following the geometry and nomenclature of Fig.~2, the spherical form of the (normalized) polarization vectors are given by

\begin{equation}
\begin{array}{ll}
\varepsilon_l^{(\pm1)}=\mp \frac{  (\cos \chi_1 \pm i \sin \chi_1) }{ \sqrt{2}}~, & 
 \varepsilon_l^{(0)}=0~, \\
\varepsilon_2^{(\pm1)}=\mp \frac{(\cos \chi_2 \cos \theta\pm i \sin \chi_2)}{\sqrt{2}},  & \varepsilon_2^{(0)}=-\cos \chi_2 \sin \theta ~, \\
\end{array}
\end{equation}
where  $\chi \equiv \chi_1$.  

The matrix elements of $r_\lambda$ can be further simplified by making use of the Wigner-Eckart theorem \cite{ros1957} and considering the overall atomic state being the product coupling of the nucleus and electron angular momenta, i.e.,

\begin{equation}
\left|\beta F m\right\rangle =\sum_{m_{I}m_{J}}\left\langle J m_{J} I m_{I} | F m \right\rangle \left|I m_{I} \right\rangle \left|\beta J m_{J}\right\rangle ~.
\end{equation}
Here, the quantities $\left\langle j_1 m_{j_1} j_2 m_{j_2} | j_3 m_{j_3} \right\rangle$ stand for the Clebsch-Gordan coefficients. 
After employing sum rules of Clebsch-Gordan coefficients \cite{ros1957}, we get 

\begin{widetext}
\begin{equation}
\left\langle \beta' F' m' J'\left|r_{\lambda}\right| \beta F m J\right\rangle = 
 (-1)^{F'+I+F+1+J'-m'}\sqrt{[F, F']}\left(\begin{array}{ccc}
F' & 1 & F\\
-m' & \lambda & m\end{array}\right)\left\{ \begin{array}{ccc}
J & I & F\\
F' & 1 & J'\end{array}\right\} \left\langle \beta' J'||r||\beta J\right\rangle ~,
\label{eq:tota_redu_jj}
\end{equation}
where the notation $[j_1, j_2,...]$ is equal to $(2j_1+1)(2j_2+1)...~$.  Since $r_\lambda$ does not act on the spin part of the wavefunction, the reduced matrix element $\left\langle J'||r_{\lambda}||J_{i}\right\rangle $ is given by 

\begin{equation}
 \left\langle \beta' J'||r||\beta J\right\rangle  = (-1)^{J' - 1/2} \sqrt{\left[J, J' \right]}\left(\begin{array}{ccc}
J' & 1 & J\\
1/2 & 0 & -1/2\end{array}\right)\left\langle n'||r||n\right\rangle  ~,
\label{eq:redu_jj}
\end{equation}
provided that $L'+L+1$ is even, where $L$ is the orbital angular momentum of the atomic state. Combining Eqs.~\eqref{eq:tota_redu_jj} and \eqref{eq:redu_jj} and rearranging the terms, the quantities $\mathcal{S}_{f \nu i}$ and $\Omega_{J_{i}J_{\nu}J_{f}}^{F_{i}F_{\nu}F_{f}}(  \theta, \chi )$ of Eqs.~(3) and (4) can be written as 
\begin{equation}
\mathcal{S}_{f \nu i}= 
\left\langle n_{f}||r||n_{\nu}\right\rangle \left\langle n_{\nu}||r||n_{i}\right\rangle = \int r^3 R_f R_\nu dr \int r^3 R_\nu R_i dr =-\frac{128 \sqrt{2}}{27 (m_{\mu n} Z)^2}~,
\label{eq:s_inte}
\end{equation}
and
\begin{equation}
\Omega_{J_{i}J_{\nu}J_{f}}^{F_{i} F_{\nu}F_{f}}(  \theta, \chi,  \eta,{\bm \varepsilon_{2}})=  [J_{\nu}, F_\nu]\sqrt{\left[F_{f},F_{i},J_{f},J_{i}\right]} \left(\begin{array}{ccc}
J_{f} & 1 & J_{\nu}\\
1/2 & 0 & -1/2\end{array}\right)\left(\begin{array}{ccc}
J_{\nu} & 1 & J_{i}\\
1/2 & 0 & -1/2\end{array}\right)  \left\{ \begin{array}{ccc}
F_{f} & 1 & F_{\nu}\\
J_{\nu} & I & J_{f}\end{array}\right\} \left\{ \begin{array}{ccc}
F_{\nu} & 1 & F_{i}\\
J_{i} & I & J_{\nu}\end{array}\right\}  \theta_{F_{f} F_{i}}^{F_{\nu}}~,
\label{eq:omega}
\end{equation}
with 
\begin{equation}
\theta_{F_{f} F_{i}}^{F_{\nu}} =  \sum_{\lambda_1, \lambda_2}   \sum_{m_{\nu}}(-1)^{\lambda_1+ \lambda_2+m_{\nu}+m_{f}+1} \hspace{0.1cm}\varepsilon_1^ {\lambda_1} \varepsilon_2^{\lambda_2 *}  \left(\begin{array}{ccc}
F_{f} & 1 & F_{\nu}\\
-m_{f} & \lambda_{2} & m_{\nu}\end{array}\right)\left(\begin{array}{ccc}
F_{\nu} & 1 & F_{i}\\
-m_{\nu} & \lambda_{1} & m_{i}\end{array}\right) ~.
\end{equation}

\end {widetext}

The functions $R$ in Eq.~\eqref{eq:s_inte} stand for the radial nonrelativistic wavefunctions, which for the case of $2s\rightarrow2p\rightarrow1s$ gives the numerical result shown on the right side of Eq.~\eqref{eq:s_inte}. The quantity $m_{\mu n}$ is the ratio between the muon-nucleus reduced mass and the electron mass. 

In case of incident linear polarized photons, the dipole radiation pattern of the scattered photon depends only on the angle $\gamma$ between the incident polarization and the direction of the scattered photon, which is related to the previous angles by $\cos \gamma= \cos \chi \sin \theta$.
 $\Lambda_{J_{i}J_{\nu}}^{F_{i} F_{\nu}}( \theta, \chi)$ and $\Xi_{J_{i}J_{\nu} J_{\nu'}}^{F_{i}F_{\nu}F_{\nu'}}( \theta, \chi)$ are parametrized in terms of this angle $\gamma$ by $a_0 + a_2 P_2(\cos\gamma)$ and $ b_2 P_2(\cos\gamma)$, respectively. The respective coefficients calculated using Eq.~\eqref{eq:omega} are listed in Table~\ref{tab:coe_para_del_Xi}.


\bibliography{HF_articles}

\begin{thebibliography}{47}%
\makeatletter
\providecommand \@ifxundefined [1]{%
 \@ifx{#1\undefined}
}%
\providecommand \@ifnum [1]{%
 \ifnum #1\expandafter \@firstoftwo
 \else \expandafter \@secondoftwo
 \fi
}%
\providecommand \@ifx [1]{%
 \ifx #1\expandafter \@firstoftwo
 \else \expandafter \@secondoftwo
 \fi
}%
\providecommand \natexlab [1]{#1}%
\providecommand \enquote  [1]{``#1''}%
\providecommand \bibnamefont  [1]{#1}%
\providecommand \bibfnamefont [1]{#1}%
\providecommand \citenamefont [1]{#1}%
\providecommand \href@noop [0]{\@secondoftwo}%
\providecommand \href [0]{\begingroup \@sanitize@url \@href}%
\providecommand \@href[1]{\@@startlink{#1}\@@href}%
\providecommand \@@href[1]{\endgroup#1\@@endlink}%
\providecommand \@sanitize@url [0]{\catcode `\\12\catcode `\$12\catcode
  `\&12\catcode `\#12\catcode `\^12\catcode `\_12\catcode `\%12\relax}%
\providecommand \@@startlink[1]{}%
\providecommand \@@endlink[0]{}%
\providecommand \url  [0]{\begingroup\@sanitize@url \@url }%
\providecommand \@url [1]{\endgroup\@href {#1}{\urlprefix }}%
\providecommand \urlprefix  [0]{URL }%
\providecommand \Eprint [0]{\href }%
\@ifxundefined \urlstyle {%
  \providecommand \doi  [0]{\begingroup \@sanitize@url \@doi}%
  \providecommand \@doi [1]{\endgroup \@@startlink {\doibase
  #1}doi:\discretionary {}{}{}#1\@@endlink }%
}{%
  \providecommand \doi  [0]{doi:\discretionary{}{}{}\begingroup
  \urlstyle{rm}\Url }%
}%
\providecommand \doibase [0]{http://dx.doi.org/}%
\providecommand \Doi [0]{\begingroup \@sanitize@url \@Doi }%
\providecommand \@Doi  [1]{\endgroup\@@startlink{\doibase#1}\@@Doi}%
\providecommand \@@Doi [1]{#1\@@endlink}%
\providecommand \selectlanguage [0]{\@gobble}%
\providecommand \bibinfo  [0]{\@secondoftwo}%
\providecommand \bibfield  [0]{\@secondoftwo}%
\providecommand \translation [1]{[#1]}%
\providecommand \BibitemOpen [0]{}%
\providecommand \bibitemStop [0]{}%
\providecommand \bibitemNoStop [0]{.\EOS\space}%
\providecommand \EOS [0]{\spacefactor3000\relax}%
\providecommand \BibitemShut  [1]{\csname bibitem#1\endcsname}%
\bibitem [{\citenamefont {Low}(1952)}]{low1952}%
  \BibitemOpen
  \bibfield  {author} {\bibinfo {author} {\bibfnamefont {F.}~\bibnamefont
  {Low}},\ }\href {http://link.aps.org/abstract/PR/v88/p53} {\bibfield
  {journal} {\bibinfo  {journal} {Phys. Rev.},\ }\textbf {\bibinfo {volume}
  {88}},\ \bibinfo {pages} {53} (\bibinfo {year} {1952})}\BibitemShut {NoStop}%
\bibitem [{\citenamefont {Labzowsky}\ \emph {et~al.}(1994)\citenamefont
  {Labzowsky}, \citenamefont {Karasiev},\ and\ \citenamefont
  {Goidenko}}]{lkg1994}%
  \BibitemOpen
  \bibfield  {author} {\bibinfo {author} {\bibfnamefont {L.}~\bibnamefont
  {Labzowsky}}, \bibinfo {author} {\bibfnamefont {V.}~\bibnamefont {Karasiev}},
  \ and\ \bibinfo {author} {\bibfnamefont {I.}~\bibnamefont {Goidenko}},\
  }\href {http://stacks.iop.org/0953-4075/27/L439} {\bibfield  {journal}
  {\bibinfo  {journal} {J. Phys. B},\ }\textbf {\bibinfo {volume} {27}},\
  \bibinfo {pages} {L439} (\bibinfo {year} {1994})}\BibitemShut {NoStop}%
\bibitem [{\citenamefont {Labzowsky}\ \emph {et~al.}(2001)\citenamefont
  {Labzowsky}, \citenamefont {Solovyev}, \citenamefont {Plunien},\ and\
  \citenamefont {Soff}}]{lsp2001}%
  \BibitemOpen
  \bibfield  {author} {\bibinfo {author} {\bibfnamefont {L.~N.}\ \bibnamefont
  {Labzowsky}}, \bibinfo {author} {\bibfnamefont {D.~A.}\ \bibnamefont
  {Solovyev}}, \bibinfo {author} {\bibfnamefont {G.}~\bibnamefont {Plunien}}, \
  and\ \bibinfo {author} {\bibfnamefont {G.}~\bibnamefont {Soff}},\ }\href
  {http://link.aps.org/doi/10.1103/PhysRevLett.87.143003} {\bibfield  {journal}
  {\bibinfo  {journal} {Phys. Rev. Lett.},\ }\textbf {\bibinfo {volume} {87}},\
  \bibinfo {pages} {143003} (\bibinfo {year} {2001})}\BibitemShut {NoStop}%
\bibitem [{\citenamefont {Labzowsky}\ \emph {et~al.}(2009)\citenamefont
  {Labzowsky}, \citenamefont {Schedrin}, \citenamefont {Solovyev},
  \citenamefont {Chernovskaya}, \citenamefont {Plunien},\ and\ \citenamefont
  {Karshenboim}}]{lga2009}%
  \BibitemOpen
  \bibfield  {author} {\bibinfo {author} {\bibfnamefont {L.}~\bibnamefont
  {Labzowsky}}, \bibinfo {author} {\bibfnamefont {G.}~\bibnamefont {Schedrin}},
  \bibinfo {author} {\bibfnamefont {D.}~\bibnamefont {Solovyev}}, \bibinfo
  {author} {\bibfnamefont {E.}~\bibnamefont {Chernovskaya}}, \bibinfo {author}
  {\bibfnamefont {G.}~\bibnamefont {Plunien}}, \ and\ \bibinfo {author}
  {\bibfnamefont {S.}~\bibnamefont {Karshenboim}},\ }\href
  {http://journals.aps.org/pra/abstract/10.1103/PhysRevA.79.052506} {\bibfield
  {journal} {\bibinfo  {journal} {Phys. Rev. A},\ }\textbf {\bibinfo {volume}
  {79}},\ \bibinfo {pages} {052506} (\bibinfo {year} {2009})}\BibitemShut
  {NoStop}%
\bibitem [{\citenamefont {Brown}\ \emph {et~al.}(2013)\citenamefont {Brown},
  \citenamefont {Wu}, \citenamefont {Porto}, \citenamefont {Sansonetti},
  \citenamefont {Simien}, \citenamefont {Brewer}, \citenamefont {Tan},\ and\
  \citenamefont {Gillaspy}}]{bup2013}%
  \BibitemOpen
  \bibfield  {author} {\bibinfo {author} {\bibfnamefont {R.~C.}\ \bibnamefont
  {Brown}}, \bibinfo {author} {\bibfnamefont {S.}~\bibnamefont {Wu}}, \bibinfo
  {author} {\bibfnamefont {J.~V.}\ \bibnamefont {Porto}}, \bibinfo {author}
  {\bibfnamefont {C.~J.}\ \bibnamefont {Sansonetti}}, \bibinfo {author}
  {\bibfnamefont {C.~E.}\ \bibnamefont {Simien}}, \bibinfo {author}
  {\bibfnamefont {S.~M.}\ \bibnamefont {Brewer}}, \bibinfo {author}
  {\bibfnamefont {J.~N.}\ \bibnamefont {Tan}}, \ and\ \bibinfo {author}
  {\bibfnamefont {J.~D.}\ \bibnamefont {Gillaspy}},\ }\href
  {http://link.aps.org/doi/10.1103/PhysRevA.87.032504} {\bibfield  {journal}
  {\bibinfo  {journal} {Phys. Rev. A},\ }\textbf {\bibinfo {volume} {87}},\
  \bibinfo {pages} {032504} (\bibinfo {year} {2013})}\BibitemShut {NoStop}%
\bibitem [{\citenamefont {Huber}\ \emph {et~al.}(1999)\citenamefont {Huber},
  \citenamefont {Gross}, \citenamefont {Weitz},\ and\ \citenamefont
  {H\"{a}nsch}}]{hgw1999}%
  \BibitemOpen
  \bibfield  {author} {\bibinfo {author} {\bibfnamefont {A.}~\bibnamefont
  {Huber}}, \bibinfo {author} {\bibfnamefont {B.}~\bibnamefont {Gross}},
  \bibinfo {author} {\bibfnamefont {M.}~\bibnamefont {Weitz}}, \ and\ \bibinfo
  {author} {\bibfnamefont {T.~W.}\ \bibnamefont {H\"{a}nsch}},\ }\href
  {http://link.aps.org/doi/10.1103/PhysRevA.59.1844} {\bibfield  {journal}
  {\bibinfo  {journal} {Phys. Rev. A},\ }\textbf {\bibinfo {volume} {59}},\
  \bibinfo {pages} {1844} (\bibinfo {year} {1999})}\BibitemShut {NoStop}%
\bibitem [{\citenamefont {Niering}\ \emph {et~al.}(2000)\citenamefont
  {Niering}, \citenamefont {Holzwarth}, \citenamefont {Reichert}, \citenamefont
  {Pokasov}, \citenamefont {Udem}, \citenamefont {Weitz}, \citenamefont
  {H\"{a}nsch}, \citenamefont {Lemonde}, \citenamefont {Santarelli},
  \citenamefont {Abgrall}, \citenamefont {Laurent}, \citenamefont {Salomon},\
  and\ \citenamefont {Clairon}}]{nhr2000}%
  \BibitemOpen
  \bibfield  {author} {\bibinfo {author} {\bibfnamefont {M.}~\bibnamefont
  {Niering}}, \bibinfo {author} {\bibfnamefont {R.}~\bibnamefont {Holzwarth}},
  \bibinfo {author} {\bibfnamefont {J.}~\bibnamefont {Reichert}}, \bibinfo
  {author} {\bibfnamefont {P.}~\bibnamefont {Pokasov}}, \bibinfo {author}
  {\bibfnamefont {T.}~\bibnamefont {Udem}}, \bibinfo {author} {\bibfnamefont
  {M.}~\bibnamefont {Weitz}}, \bibinfo {author} {\bibfnamefont {T.~W.}\
  \bibnamefont {H\"{a}nsch}}, \bibinfo {author} {\bibfnamefont
  {P.}~\bibnamefont {Lemonde}}, \bibinfo {author} {\bibfnamefont
  {G.}~\bibnamefont {Santarelli}}, \bibinfo {author} {\bibfnamefont
  {M.}~\bibnamefont {Abgrall}}, \bibinfo {author} {\bibfnamefont
  {P.}~\bibnamefont {Laurent}}, \bibinfo {author} {\bibfnamefont
  {C.}~\bibnamefont {Salomon}}, \ and\ \bibinfo {author} {\bibfnamefont
  {A.}~\bibnamefont {Clairon}},\ }\href
  {http://link.aps.org/doi/10.1103/PhysRevLett.84.5496} {\bibfield  {journal}
  {\bibinfo  {journal} {Phys. Rev. Lett.},\ }\textbf {\bibinfo {volume} {84}},\
  \bibinfo {pages} {5496} (\bibinfo {year} {2000})}\BibitemShut {NoStop}%
\bibitem [{\citenamefont {Parthey}\ \emph {et~al.}(2011)\citenamefont
  {Parthey}, \citenamefont {Matveev}, \citenamefont {Alnis}, \citenamefont
  {Bernhardt}, \citenamefont {Beyer}, \citenamefont {Holzwarth}, \citenamefont
  {Maistrou}, \citenamefont {Pohl}, \citenamefont {Predehl}, \citenamefont
  {Udem}, \citenamefont {Wilken}, \citenamefont {Kolachevsky}, \citenamefont
  {Abgrall}, \citenamefont {Rovera}, \citenamefont {Salomon}, \citenamefont
  {Laurent},\ and\ \citenamefont {H\"{a}nsch}}]{pma2011}%
  \BibitemOpen
  \bibfield  {author} {\bibinfo {author} {\bibfnamefont {C.~G.}\ \bibnamefont
  {Parthey}}, \bibinfo {author} {\bibfnamefont {A.}~\bibnamefont {Matveev}},
  \bibinfo {author} {\bibfnamefont {J.}~\bibnamefont {Alnis}}, \bibinfo
  {author} {\bibfnamefont {B.}~\bibnamefont {Bernhardt}}, \bibinfo {author}
  {\bibfnamefont {A.}~\bibnamefont {Beyer}}, \bibinfo {author} {\bibfnamefont
  {R.}~\bibnamefont {Holzwarth}}, \bibinfo {author} {\bibfnamefont
  {A.}~\bibnamefont {Maistrou}}, \bibinfo {author} {\bibfnamefont
  {R.}~\bibnamefont {Pohl}}, \bibinfo {author} {\bibfnamefont {K.}~\bibnamefont
  {Predehl}}, \bibinfo {author} {\bibfnamefont {T.}~\bibnamefont {Udem}},
  \bibinfo {author} {\bibfnamefont {T.}~\bibnamefont {Wilken}}, \bibinfo
  {author} {\bibfnamefont {N.}~\bibnamefont {Kolachevsky}}, \bibinfo {author}
  {\bibfnamefont {M.}~\bibnamefont {Abgrall}}, \bibinfo {author} {\bibfnamefont
  {D.}~\bibnamefont {Rovera}}, \bibinfo {author} {\bibfnamefont
  {C.}~\bibnamefont {Salomon}}, \bibinfo {author} {\bibfnamefont
  {P.}~\bibnamefont {Laurent}}, \ and\ \bibinfo {author} {\bibfnamefont
  {T.~W.}\ \bibnamefont {H\"{a}nsch}},\ }\href
  {http://link.aps.org/doi/10.1103/PhysRevLett.107.203001} {\bibfield
  {journal} {\bibinfo  {journal} {Phys. Rev. Lett.},\ }\textbf {\bibinfo
  {volume} {107}},\ \bibinfo {pages} {203001} (\bibinfo {year}
  {2011})}\BibitemShut {NoStop}%
\bibitem [{\citenamefont {Jentschura}\ and\ \citenamefont
  {Mohr}(2002)}]{jem2002}%
  \BibitemOpen
  \bibfield  {author} {\bibinfo {author} {\bibfnamefont {U.~D.}\ \bibnamefont
  {Jentschura}}\ and\ \bibinfo {author} {\bibfnamefont {P.~J.}\ \bibnamefont
  {Mohr}},\ }\href {http://dx.doi.org/10.1139/p02-019} {\bibfield  {journal}
  {\bibinfo  {journal} {Can. J. Phys.},\ }\textbf {\bibinfo {volume} {80}},\
  \bibinfo {pages} {633} (\bibinfo {year} {2002})}\BibitemShut {NoStop}%
\bibitem [{\citenamefont {Labzowsky}\ \emph {et~al.}(2007)\citenamefont
  {Labzowsky}, \citenamefont {Schedrin}, \citenamefont {Solovyev},\ and\
  \citenamefont {Plunien}}]{lss2007}%
  \BibitemOpen
  \bibfield  {author} {\bibinfo {author} {\bibfnamefont {L.~N.}\ \bibnamefont
  {Labzowsky}}, \bibinfo {author} {\bibfnamefont {G.}~\bibnamefont {Schedrin}},
  \bibinfo {author} {\bibfnamefont {D.}~\bibnamefont {Solovyev}}, \ and\
  \bibinfo {author} {\bibfnamefont {G.}~\bibnamefont {Plunien}},\ }\href
  {http://www.nrcresearchpress.com/doi/abs/10.1139/p07-014} {\bibfield
  {journal} {\bibinfo  {journal} {Can. J. Phys.},\ }\textbf {\bibinfo {volume}
  {85}},\ \bibinfo {pages} {585} (\bibinfo {year} {2007})}\BibitemShut
  {NoStop}%
\bibitem [{\citenamefont {Karshenboim}\ and\ \citenamefont
  {Ivanov}(2008)}]{kiv2008}%
  \BibitemOpen
  \bibfield  {author} {\bibinfo {author} {\bibfnamefont {S.~G.}\ \bibnamefont
  {Karshenboim}}\ and\ \bibinfo {author} {\bibfnamefont {V.~G.}\ \bibnamefont
  {Ivanov}},\ }\href {http://dx.doi.org/10.1134/S1063773708050010} {\bibfield
  {journal} {\bibinfo  {journal} {Astron. Lett.},\ }\textbf {\bibinfo {volume}
  {34}},\ \bibinfo {pages} {289} (\bibinfo {year} {2008})}\BibitemShut
  {NoStop}%
\bibitem [{\citenamefont {Jentschura}\ \emph {et~al.}(2003)\citenamefont
  {Jentschura}, \citenamefont {Evers}, \citenamefont {Haas},\ and\
  \citenamefont {Keitel}}]{jue2003}%
  \BibitemOpen
  \bibfield  {author} {\bibinfo {author} {\bibfnamefont {U.~D.}\ \bibnamefont
  {Jentschura}}, \bibinfo {author} {\bibfnamefont {J.}~\bibnamefont {Evers}},
  \bibinfo {author} {\bibfnamefont {M.}~\bibnamefont {Haas}}, \ and\ \bibinfo
  {author} {\bibfnamefont {C.~H.}\ \bibnamefont {Keitel}},\ }\href
  {http://link.aps.org/doi/10.1103/PhysRevLett.91.253601} {\bibfield  {journal}
  {\bibinfo  {journal} {Phys. Rev. Lett.},\ }\textbf {\bibinfo {volume} {91}},\
  \bibinfo {pages} {253601} (\bibinfo {year} {2003})}\BibitemShut {NoStop}%
\bibitem [{\citenamefont {Franken}(1961)}]{fra1961}%
  \BibitemOpen
  \bibfield  {author} {\bibinfo {author} {\bibfnamefont {P.~A.}\ \bibnamefont
  {Franken}},\ }\href {http://link.aps.org/doi/10.1103/PhysRev.121.508}
  {\bibfield  {journal} {\bibinfo  {journal} {Phys. Rev.},\ }\textbf {\bibinfo
  {volume} {121}},\ \bibinfo {pages} {508} (\bibinfo {year}
  {1961})}\BibitemShut {NoStop}%
\bibitem [{\citenamefont {Walkup}\ \emph {et~al.}(1982)\citenamefont {Walkup},
  \citenamefont {Migdall},\ and\ \citenamefont {Pritchard}}]{wmp1982}%
  \BibitemOpen
  \bibfield  {author} {\bibinfo {author} {\bibfnamefont {R.}~\bibnamefont
  {Walkup}}, \bibinfo {author} {\bibfnamefont {A.~L.}\ \bibnamefont {Migdall}},
  \ and\ \bibinfo {author} {\bibfnamefont {D.~E.}\ \bibnamefont {Pritchard}},\
  }\href {http://link.aps.org/doi/10.1103/PhysRevA.25.3114} {\bibfield
  {journal} {\bibinfo  {journal} {Phys. Rev. A},\ }\textbf {\bibinfo {volume}
  {25}},\ \bibinfo {pages} {3114} (\bibinfo {year} {1982})}\BibitemShut
  {NoStop}%
\bibitem [{\citenamefont {Sansonetti}\ \emph {et~al.}(2011)\citenamefont
  {Sansonetti}, \citenamefont {Simien}, \citenamefont {Gillaspy}, \citenamefont
  {Tan}, \citenamefont {Brewer}, \citenamefont {Brown}, \citenamefont {Wu},\
  and\ \citenamefont {Porto}}]{scs2011}%
  \BibitemOpen
  \bibfield  {author} {\bibinfo {author} {\bibfnamefont {C.~J.}\ \bibnamefont
  {Sansonetti}}, \bibinfo {author} {\bibfnamefont {C.~E.}\ \bibnamefont
  {Simien}}, \bibinfo {author} {\bibfnamefont {J.~D.}\ \bibnamefont
  {Gillaspy}}, \bibinfo {author} {\bibfnamefont {J.~N.}\ \bibnamefont {Tan}},
  \bibinfo {author} {\bibfnamefont {S.~M.}\ \bibnamefont {Brewer}}, \bibinfo
  {author} {\bibfnamefont {R.~C.}\ \bibnamefont {Brown}}, \bibinfo {author}
  {\bibfnamefont {S.}~\bibnamefont {Wu}}, \ and\ \bibinfo {author}
  {\bibfnamefont {J.~V.}\ \bibnamefont {Porto}},\ }\href
  {http://link.aps.org/doi/10.1103/PhysRevLett.107.023001} {\bibfield
  {journal} {\bibinfo  {journal} {Phys. Rev. Lett.},\ }\textbf {\bibinfo
  {volume} {107}},\ \bibinfo {pages} {023001} (\bibinfo {year}
  {2011})}\BibitemShut {NoStop}%
\bibitem [{\citenamefont {Horbatsch}\ and\ \citenamefont
  {Hessels}(2010)}]{hoh2010}%
  \BibitemOpen
  \bibfield  {author} {\bibinfo {author} {\bibfnamefont {M.}~\bibnamefont
  {Horbatsch}}\ and\ \bibinfo {author} {\bibfnamefont {E.~A.}\ \bibnamefont
  {Hessels}},\ }\href {http://link.aps.org/doi/10.1103/PhysRevA.82.052519}
  {\bibfield  {journal} {\bibinfo  {journal} {Phys. Rev. A},\ }\textbf
  {\bibinfo {volume} {82}},\ \bibinfo {pages} {052519} (\bibinfo {year}
  {2010})}\BibitemShut {NoStop}%
\bibitem [{\citenamefont {Marsman}\ \emph
  {et~al.}(2012){\natexlab{a}}\citenamefont {Marsman}, \citenamefont
  {Horbatsch},\ and\ \citenamefont {Hessels}}]{mhh2012}%
  \BibitemOpen
  \bibfield  {author} {\bibinfo {author} {\bibfnamefont {A.}~\bibnamefont
  {Marsman}}, \bibinfo {author} {\bibfnamefont {M.}~\bibnamefont {Horbatsch}},
  \ and\ \bibinfo {author} {\bibfnamefont {E.~A.}\ \bibnamefont {Hessels}},\
  }\href {http://link.aps.org/doi/10.1103/PhysRevA.86.012510} {\bibfield
  {journal} {\bibinfo  {journal} {Phys. Rev. A},\ }\textbf {\bibinfo {volume}
  {86}},\ \bibinfo {pages} {012510} (\bibinfo {year}
  {2012}{\natexlab{a}})}\BibitemShut {NoStop}%
\bibitem [{\citenamefont {Marsman}\ \emph
  {et~al.}(2012){\natexlab{b}}\citenamefont {Marsman}, \citenamefont
  {Horbatsch},\ and\ \citenamefont {Hessels}}]{mhh2012b}%
  \BibitemOpen
  \bibfield  {author} {\bibinfo {author} {\bibfnamefont {A.}~\bibnamefont
  {Marsman}}, \bibinfo {author} {\bibfnamefont {M.}~\bibnamefont {Horbatsch}},
  \ and\ \bibinfo {author} {\bibfnamefont {E.~A.}\ \bibnamefont {Hessels}},\
  }\href {http://link.aps.org/doi/10.1103/PhysRevA.86.040501} {\bibfield
  {journal} {\bibinfo  {journal} {Phys. Rev. A},\ }\textbf {\bibinfo {volume}
  {86}},\ \bibinfo {pages} {040501} (\bibinfo {year}
  {2012}{\natexlab{b}})}\BibitemShut {NoStop}%
\bibitem [{\citenamefont {Marsman}\ \emph {et~al.}(2014)\citenamefont
  {Marsman}, \citenamefont {Hessels},\ and\ \citenamefont
  {Horbatsch}}]{mhh2014}%
  \BibitemOpen
  \bibfield  {author} {\bibinfo {author} {\bibfnamefont {A.}~\bibnamefont
  {Marsman}}, \bibinfo {author} {\bibfnamefont {E.~A.}\ \bibnamefont
  {Hessels}}, \ and\ \bibinfo {author} {\bibfnamefont {M.}~\bibnamefont
  {Horbatsch}},\ }\href {http://link.aps.org/doi/10.1103/PhysRevA.89.043403}
  {\bibfield  {journal} {\bibinfo  {journal} {Phys. Rev. A},\ }\textbf
  {\bibinfo {volume} {89}},\ \bibinfo {pages} {043403} (\bibinfo {year}
  {2014})}\BibitemShut {NoStop}%
\bibitem [{\citenamefont {Marsman}\ \emph
  {et~al.}(2015){\natexlab{a}}\citenamefont {Marsman}, \citenamefont
  {Horbatsch},\ and\ \citenamefont {Hessels}}]{mhh2015}%
  \BibitemOpen
  \bibfield  {author} {\bibinfo {author} {\bibfnamefont {A.}~\bibnamefont
  {Marsman}}, \bibinfo {author} {\bibfnamefont {M.}~\bibnamefont {Horbatsch}},
  \ and\ \bibinfo {author} {\bibfnamefont {E.~A.}\ \bibnamefont {Hessels}},\
  }\href {http://link.aps.org/doi/10.1103/PhysRevA.91.062506} {\bibfield
  {journal} {\bibinfo  {journal} {Phys. Rev. A},\ }\textbf {\bibinfo {volume}
  {91}},\ \bibinfo {pages} {062506} (\bibinfo {year}
  {2015}{\natexlab{a}})}\BibitemShut {NoStop}%
\bibitem [{\citenamefont {Marsman}\ \emph
  {et~al.}(2015){\natexlab{b}}\citenamefont {Marsman}, \citenamefont
  {Horbatsch},\ and\ \citenamefont {Hessels}}]{mhhb2015}%
  \BibitemOpen
  \bibfield  {author} {\bibinfo {author} {\bibfnamefont {A.}~\bibnamefont
  {Marsman}}, \bibinfo {author} {\bibfnamefont {M.}~\bibnamefont {Horbatsch}},
  \ and\ \bibinfo {author} {\bibfnamefont {E.~A.}\ \bibnamefont {Hessels}},\
  }\href
  {http://scitation.aip.org/content/aip/journal/jpcrd/44/3/10.1063/1.4922796}
  {\bibfield  {journal} {\bibinfo  {journal} {Journal of Physical and Chemical
  Reference Data},\ }\textbf {\bibinfo {volume} {44}},\ \bibinfo {pages}
  {031207} (\bibinfo {year} {2015}{\natexlab{b}})}\BibitemShut {NoStop}%
\bibitem [{\citenamefont {Yost}\ \emph {et~al.}(2014)\citenamefont {Yost},
  \citenamefont {Matveev}, \citenamefont {Peters}, \citenamefont {Beyer},
  \citenamefont {H\"{a}nsch},\ and\ \citenamefont {Udem}}]{mpb2014}%
  \BibitemOpen
  \bibfield  {author} {\bibinfo {author} {\bibfnamefont {D.~C.}\ \bibnamefont
  {Yost}}, \bibinfo {author} {\bibfnamefont {A.}~\bibnamefont {Matveev}},
  \bibinfo {author} {\bibfnamefont {E.}~\bibnamefont {Peters}}, \bibinfo
  {author} {\bibfnamefont {A.}~\bibnamefont {Beyer}}, \bibinfo {author}
  {\bibfnamefont {T.~W.}\ \bibnamefont {H\"{a}nsch}}, \ and\ \bibinfo {author}
  {\bibfnamefont {T.}~\bibnamefont {Udem}},\ }\href
  {http://link.aps.org/doi/10.1103/PhysRevA.90.012512} {\bibfield  {journal}
  {\bibinfo  {journal} {Phys. Rev. A},\ }\textbf {\bibinfo {volume} {90}},\
  \bibinfo {pages} {012512} (\bibinfo {year} {2014})}\BibitemShut {NoStop}%
\bibitem [{\citenamefont {Borie}(2012)}]{bor2012}%
  \BibitemOpen
  \bibfield  {author} {\bibinfo {author} {\bibfnamefont {E.}~\bibnamefont
  {Borie}},\ }\href
  {http://www.sciencedirect.com/science/article/pii/S0003491611001904}
  {\bibfield  {journal} {\bibinfo  {journal} {Ann. Phys.},\ }\textbf {\bibinfo
  {volume} {327}},\ \bibinfo {pages} {733} (\bibinfo {year}
  {2012})}\BibitemShut {NoStop}%
\bibitem [{\citenamefont {Antognini}\ \emph
  {et~al.}(2013){\natexlab{a}}\citenamefont {Antognini}, \citenamefont
  {Kottmann}, \citenamefont {Biraben}, \citenamefont {Indelicato},
  \citenamefont {Nez},\ and\ \citenamefont {Pohl}}]{akb2013}%
  \BibitemOpen
  \bibfield  {author} {\bibinfo {author} {\bibfnamefont {A.}~\bibnamefont
  {Antognini}}, \bibinfo {author} {\bibfnamefont {F.}~\bibnamefont {Kottmann}},
  \bibinfo {author} {\bibfnamefont {F.}~\bibnamefont {Biraben}}, \bibinfo
  {author} {\bibfnamefont {P.}~\bibnamefont {Indelicato}}, \bibinfo {author}
  {\bibfnamefont {F.}~\bibnamefont {Nez}}, \ and\ \bibinfo {author}
  {\bibfnamefont {R.}~\bibnamefont {Pohl}},\ }\href
  {http://www.sciencedirect.com/science/article/pii/S0003491612002102}
  {\bibfield  {journal} {\bibinfo  {journal} {Ann. Phys.},\ }\textbf {\bibinfo
  {volume} {331}},\ \bibinfo {pages} {127} (\bibinfo {year}
  {2013}{\natexlab{a}})}\BibitemShut {NoStop}%
\bibitem [{\citenamefont {Milotti}(1998)}]{med1998}%
  \BibitemOpen
  \bibfield  {author} {\bibinfo {author} {\bibfnamefont {E.}~\bibnamefont
  {Milotti}},\ }\href
  {http://www.sciencedirect.com/science/article/pii/S0092640X98907902}
  {\bibfield  {journal} {\bibinfo  {journal} {At. Data Nucl. Data Tables},\
  }\textbf {\bibinfo {volume} {70}},\ \bibinfo {pages} {137} (\bibinfo {year}
  {1998})}\BibitemShut {NoStop}%
\bibitem [{\citenamefont {Safari}\ \emph {et~al.}(2012)\citenamefont {Safari},
  \citenamefont {Amaro}, \citenamefont {Fritzsche}, \citenamefont {Santos},
  \citenamefont {Tashenov},\ and\ \citenamefont {Fratini}}]{saf2012b}%
  \BibitemOpen
  \bibfield  {author} {\bibinfo {author} {\bibfnamefont {L.}~\bibnamefont
  {Safari}}, \bibinfo {author} {\bibfnamefont {P.}~\bibnamefont {Amaro}},
  \bibinfo {author} {\bibfnamefont {S.}~\bibnamefont {Fritzsche}}, \bibinfo
  {author} {\bibfnamefont {J.~P.}\ \bibnamefont {Santos}}, \bibinfo {author}
  {\bibfnamefont {S.}~\bibnamefont {Tashenov}}, \ and\ \bibinfo {author}
  {\bibfnamefont {F.}~\bibnamefont {Fratini}},\ }\href
  {http://link.aps.org/doi/10.1103/PhysRevA.86.043405} {\bibfield  {journal}
  {\bibinfo  {journal} {Phys. Rev. A},\ }\textbf {\bibinfo {volume} {86}},\
  \bibinfo {pages} {043405} (\bibinfo {year} {2012})}\BibitemShut {NoStop}%
\bibitem [{\citenamefont {Safari}\ \emph {et~al.}(2015)\citenamefont {Safari},
  \citenamefont {Amaro}, \citenamefont {Santos},\ and\ \citenamefont
  {Fratini}}]{sas2015}%
  \BibitemOpen
  \bibfield  {author} {\bibinfo {author} {\bibfnamefont {L.}~\bibnamefont
  {Safari}}, \bibinfo {author} {\bibfnamefont {P.}~\bibnamefont {Amaro}},
  \bibinfo {author} {\bibfnamefont {J.~P.}\ \bibnamefont {Santos}}, \ and\
  \bibinfo {author} {\bibfnamefont {F.}~\bibnamefont {Fratini}},\ }\href
  {http://www.sciencedirect.com/science/article/pii/S0969806X14003351}
  {\bibfield  {journal} {\bibinfo  {journal} {Radiation Physics and
  Chemistry},\ }\textbf {\bibinfo {volume} {106}},\ \bibinfo {pages} {271}
  (\bibinfo {year} {2015})}\BibitemShut {NoStop}%
\bibitem [{\citenamefont {Martynenko}(2005)}]{mar2005}%
  \BibitemOpen
  \bibfield  {author} {\bibinfo {author} {\bibfnamefont {A.~P.}\ \bibnamefont
  {Martynenko}},\ }\href {http://link.aps.org/abstract/PRA/v71/e022506}
  {\bibfield  {journal} {\bibinfo  {journal} {Phys. Rev. A},\ }\textbf
  {\bibinfo {volume} {71}},\ \bibinfo {pages} {022506} (\bibinfo {year}
  {2005})}\BibitemShut {NoStop}%
\bibitem [{\citenamefont {Martynenko}(2008)}]{mar2008b}%
  \BibitemOpen
  \bibfield  {author} {\bibinfo {author} {\bibfnamefont {A.}~\bibnamefont
  {Martynenko}},\ }\href {http://dx.doi.org/10.1134/S1063776108040079}
  {\bibfield  {journal} {\bibinfo  {journal} {JETP},\ }\textbf {\bibinfo
  {volume} {106}},\ \bibinfo {pages} {690} (\bibinfo {year}
  {2008})}\BibitemShut {NoStop}%
\bibitem [{\citenamefont {Borie}(2005)}]{bor2005}%
  \BibitemOpen
  \bibfield  {author} {\bibinfo {author} {\bibfnamefont {E.}~\bibnamefont
  {Borie}},\ }\href {http://link.aps.org/abstract/PRA/v71/e032508} {\bibfield
  {journal} {\bibinfo  {journal} {Phys. Rev. A},\ }\textbf {\bibinfo {volume}
  {71}},\ \bibinfo {pages} {032508} (\bibinfo {year} {2005})}\BibitemShut
  {NoStop}%
\bibitem [{\citenamefont {Pohl}\ \emph {et~al.}(2010)\citenamefont {Pohl},
  \citenamefont {Antognini}, \citenamefont {Nez}, \citenamefont {Amaro},
  \citenamefont {Biraben}, \citenamefont {Cardoso}, \citenamefont {Covita},
  \citenamefont {Dax}, \citenamefont {Dhawan}, \citenamefont {Fernandes},
  \citenamefont {Giesen}, \citenamefont {Graf}, \citenamefont {H\"{a}nsch},
  \citenamefont {Indelicato}, \citenamefont {Julien}, \citenamefont {Kao},
  \citenamefont {Knowles}, \citenamefont {Le~Bigot}, \citenamefont {Liu},
  \citenamefont {Lopes}, \citenamefont {Ludhova}, \citenamefont {Monteiro},
  \citenamefont {Mulhauser}, \citenamefont {Nebel}, \citenamefont {Rabinowitz},
  \citenamefont {dos Santos}, \citenamefont {Schaller}, \citenamefont
  {Schuhmann}, \citenamefont {Schwob}, \citenamefont {Taqqu}, \citenamefont
  {Veloso},\ and\ \citenamefont {Kottmann}}]{pan2010}%
  \BibitemOpen
  \bibfield  {author} {\bibinfo {author} {\bibfnamefont {R.}~\bibnamefont
  {Pohl}}, \bibinfo {author} {\bibfnamefont {A.}~\bibnamefont {Antognini}},
  \bibinfo {author} {\bibfnamefont {F.}~\bibnamefont {Nez}}, \bibinfo {author}
  {\bibfnamefont {F.~D.}\ \bibnamefont {Amaro}}, \bibinfo {author}
  {\bibfnamefont {F.}~\bibnamefont {Biraben}}, \bibinfo {author} {\bibfnamefont
  {J.~M.~R.}\ \bibnamefont {Cardoso}}, \bibinfo {author} {\bibfnamefont
  {D.~S.}\ \bibnamefont {Covita}}, \bibinfo {author} {\bibfnamefont
  {A.}~\bibnamefont {Dax}}, \bibinfo {author} {\bibfnamefont {S.}~\bibnamefont
  {Dhawan}}, \bibinfo {author} {\bibfnamefont {L.~M.~P.}\ \bibnamefont
  {Fernandes}}, \bibinfo {author} {\bibfnamefont {A.}~\bibnamefont {Giesen}},
  \bibinfo {author} {\bibfnamefont {T.}~\bibnamefont {Graf}}, \bibinfo {author}
  {\bibfnamefont {T.~W.}\ \bibnamefont {H\"{a}nsch}}, \bibinfo {author}
  {\bibfnamefont {P.}~\bibnamefont {Indelicato}}, \bibinfo {author}
  {\bibfnamefont {L.}~\bibnamefont {Julien}}, \bibinfo {author} {\bibfnamefont
  {C.-Y.}\ \bibnamefont {Kao}}, \bibinfo {author} {\bibfnamefont
  {P.}~\bibnamefont {Knowles}}, \bibinfo {author} {\bibfnamefont {E.-O.}\
  \bibnamefont {Le~Bigot}}, \bibinfo {author} {\bibfnamefont {Y.-W.}\
  \bibnamefont {Liu}}, \bibinfo {author} {\bibfnamefont {J.~A.~M.}\
  \bibnamefont {Lopes}}, \bibinfo {author} {\bibfnamefont {L.}~\bibnamefont
  {Ludhova}}, \bibinfo {author} {\bibfnamefont {C.~M.~B.}\ \bibnamefont
  {Monteiro}}, \bibinfo {author} {\bibfnamefont {F.}~\bibnamefont {Mulhauser}},
  \bibinfo {author} {\bibfnamefont {T.}~\bibnamefont {Nebel}}, \bibinfo
  {author} {\bibfnamefont {P.}~\bibnamefont {Rabinowitz}}, \bibinfo {author}
  {\bibfnamefont {J.~M.~F.}\ \bibnamefont {dos Santos}}, \bibinfo {author}
  {\bibfnamefont {L.~A.}\ \bibnamefont {Schaller}}, \bibinfo {author}
  {\bibfnamefont {K.}~\bibnamefont {Schuhmann}}, \bibinfo {author}
  {\bibfnamefont {C.}~\bibnamefont {Schwob}}, \bibinfo {author} {\bibfnamefont
  {D.}~\bibnamefont {Taqqu}}, \bibinfo {author} {\bibfnamefont {J.~F. C.~A.}\
  \bibnamefont {Veloso}}, \ and\ \bibinfo {author} {\bibfnamefont
  {F.}~\bibnamefont {Kottmann}},\ }\href
  {http://dx.doi.org/10.1038/nature09250} {\bibfield  {journal} {\bibinfo
  {journal} {Nature},\ }\textbf {\bibinfo {volume} {466}},\ \bibinfo {pages}
  {213} (\bibinfo {year} {2010})}\BibitemShut {NoStop}%
\bibitem [{\citenamefont {Antognini}\ \emph
  {et~al.}(2013){\natexlab{b}}\citenamefont {Antognini}, \citenamefont {Nez},
  \citenamefont {Schuhmann}, \citenamefont {Amaro}, \citenamefont {Biraben},
  \citenamefont {Cardoso}, \citenamefont {Covita}, \citenamefont {Dax},
  \citenamefont {Dhawan}, \citenamefont {Diepold}, \citenamefont {Fernandes},
  \citenamefont {Giesen}, \citenamefont {Gouvea}, \citenamefont {Graf},
  \citenamefont {H\"{a}nsch}, \citenamefont {Indelicato}, \citenamefont
  {Julien}, \citenamefont {Kao}, \citenamefont {Knowles}, \citenamefont
  {Kottmann}, \citenamefont {Le~Bigot}, \citenamefont {Liu}, \citenamefont
  {Lopes}, \citenamefont {Ludhova}, \citenamefont {Monteiro}, \citenamefont
  {Mulhauser}, \citenamefont {Nebel}, \citenamefont {Rabinowitz}, \citenamefont
  {dos Santos}, \citenamefont {Schaller}, \citenamefont {Schwob}, \citenamefont
  {Taqqu}, \citenamefont {Veloso}, \citenamefont {Vogelsang},\ and\
  \citenamefont {Pohl}}]{ans2013}%
  \BibitemOpen
  \bibfield  {author} {\bibinfo {author} {\bibfnamefont {A.}~\bibnamefont
  {Antognini}}, \bibinfo {author} {\bibfnamefont {F.}~\bibnamefont {Nez}},
  \bibinfo {author} {\bibfnamefont {K.}~\bibnamefont {Schuhmann}}, \bibinfo
  {author} {\bibfnamefont {F.~D.}\ \bibnamefont {Amaro}}, \bibinfo {author}
  {\bibfnamefont {F.}~\bibnamefont {Biraben}}, \bibinfo {author} {\bibfnamefont
  {J.~M.~R.}\ \bibnamefont {Cardoso}}, \bibinfo {author} {\bibfnamefont
  {D.~S.}\ \bibnamefont {Covita}}, \bibinfo {author} {\bibfnamefont
  {A.}~\bibnamefont {Dax}}, \bibinfo {author} {\bibfnamefont {S.}~\bibnamefont
  {Dhawan}}, \bibinfo {author} {\bibfnamefont {M.}~\bibnamefont {Diepold}},
  \bibinfo {author} {\bibfnamefont {L.~M.~P.}\ \bibnamefont {Fernandes}},
  \bibinfo {author} {\bibfnamefont {A.}~\bibnamefont {Giesen}}, \bibinfo
  {author} {\bibfnamefont {A.~L.}\ \bibnamefont {Gouvea}}, \bibinfo {author}
  {\bibfnamefont {T.}~\bibnamefont {Graf}}, \bibinfo {author} {\bibfnamefont
  {T.~W.}\ \bibnamefont {H\"{a}nsch}}, \bibinfo {author} {\bibfnamefont
  {P.}~\bibnamefont {Indelicato}}, \bibinfo {author} {\bibfnamefont
  {L.}~\bibnamefont {Julien}}, \bibinfo {author} {\bibfnamefont {C.-Y.}\
  \bibnamefont {Kao}}, \bibinfo {author} {\bibfnamefont {P.}~\bibnamefont
  {Knowles}}, \bibinfo {author} {\bibfnamefont {F.}~\bibnamefont {Kottmann}},
  \bibinfo {author} {\bibfnamefont {E.-O.}\ \bibnamefont {Le~Bigot}}, \bibinfo
  {author} {\bibfnamefont {Y.-W.}\ \bibnamefont {Liu}}, \bibinfo {author}
  {\bibfnamefont {J.~A.~M.}\ \bibnamefont {Lopes}}, \bibinfo {author}
  {\bibfnamefont {L.}~\bibnamefont {Ludhova}}, \bibinfo {author} {\bibfnamefont
  {C.~M.~B.}\ \bibnamefont {Monteiro}}, \bibinfo {author} {\bibfnamefont
  {F.}~\bibnamefont {Mulhauser}}, \bibinfo {author} {\bibfnamefont
  {T.}~\bibnamefont {Nebel}}, \bibinfo {author} {\bibfnamefont
  {P.}~\bibnamefont {Rabinowitz}}, \bibinfo {author} {\bibfnamefont {J.~M.~F.}\
  \bibnamefont {dos Santos}}, \bibinfo {author} {\bibfnamefont {L.~A.}\
  \bibnamefont {Schaller}}, \bibinfo {author} {\bibfnamefont {C.}~\bibnamefont
  {Schwob}}, \bibinfo {author} {\bibfnamefont {D.}~\bibnamefont {Taqqu}},
  \bibinfo {author} {\bibfnamefont {J.~F. C.~A.}\ \bibnamefont {Veloso}},
  \bibinfo {author} {\bibfnamefont {J.}~\bibnamefont {Vogelsang}}, \ and\
  \bibinfo {author} {\bibfnamefont {R.}~\bibnamefont {Pohl}},\ }\href
  {http://www.sciencemag.org/content/339/6118/417.abstractN2 - Accurate
  knowledge of the charge and Zemach radii of the proton is essential, not only
  for understanding its structure but also as input for tests of bound-state
  quantum electrodynamics and its predictions for the energy levels of
  hydrogen. These radii may be extracted from the laser spectroscopy of muonic
  hydrogen (μp, that is, a proton orbited by a muon). We measured the
  transition frequency in μp to be 54611.16(1.05) gigahertz (numbers in
  parentheses indicate one standard deviation of uncertainty) and reevaluated
  the transition frequency, yielding 49881.35(65) gigahertz. From the
  measurements, we determined the Zemach radius, rZ = 1.082(37) femtometers,
  and the magnetic radius, rM = 0.87(6) femtometer, of the proton. We also
  extracted the charge radius, rE = 0.84087(39) femtometer, with an order of
  magnitude more precision than the 2010-CODATA value and at 7σ variance with
  respect to it, thus reinforcing the proton radius puzzle.} {\bibfield
  {journal} {\bibinfo  {journal} {Science},\ }\textbf {\bibinfo {volume}
  {339}},\ \bibinfo {pages} {417} (\bibinfo {year}
  {2013}{\natexlab{b}})}\BibitemShut {NoStop}%
\bibitem [{\citenamefont {Antognini}\ \emph {et~al.}(2011)\citenamefont
  {Antognini}, \citenamefont {Biraben}, \citenamefont {Cardoso}, \citenamefont
  {Covita}, \citenamefont {Dax}, \citenamefont {Fernandes}, \citenamefont
  {Gouvea}, \citenamefont {Graf}, \citenamefont {H\"{a}nsch}, \citenamefont
  {Hildebrandt}, \citenamefont {Indelicato}, \citenamefont {Julien},
  \citenamefont {Kirch}, \citenamefont {Kottmann}, \citenamefont {Liu},
  \citenamefont {Monteiro}, \citenamefont {Mulhauser}, \citenamefont {Nebel},
  \citenamefont {Nez}, \citenamefont {dos Santos}, \citenamefont {Schuhmann},
  \citenamefont {Taqqu}, \citenamefont {Veloso}, \citenamefont {Voss},\ and\
  \citenamefont {Pohl}}]{abc2011}%
  \BibitemOpen
  \bibfield  {author} {\bibinfo {author} {\bibfnamefont {A.}~\bibnamefont
  {Antognini}}, \bibinfo {author} {\bibfnamefont {F.}~\bibnamefont {Biraben}},
  \bibinfo {author} {\bibfnamefont {J.~M.~R.}\ \bibnamefont {Cardoso}},
  \bibinfo {author} {\bibfnamefont {D.~S.}\ \bibnamefont {Covita}}, \bibinfo
  {author} {\bibfnamefont {A.}~\bibnamefont {Dax}}, \bibinfo {author}
  {\bibfnamefont {L.~M.~P.}\ \bibnamefont {Fernandes}}, \bibinfo {author}
  {\bibfnamefont {A.~L.}\ \bibnamefont {Gouvea}}, \bibinfo {author}
  {\bibfnamefont {T.}~\bibnamefont {Graf}}, \bibinfo {author} {\bibfnamefont
  {T.~W.}\ \bibnamefont {H\"{a}nsch}}, \bibinfo {author} {\bibfnamefont
  {M.}~\bibnamefont {Hildebrandt}}, \bibinfo {author} {\bibfnamefont
  {P.}~\bibnamefont {Indelicato}}, \bibinfo {author} {\bibfnamefont
  {L.}~\bibnamefont {Julien}}, \bibinfo {author} {\bibfnamefont
  {K.}~\bibnamefont {Kirch}}, \bibinfo {author} {\bibfnamefont
  {F.}~\bibnamefont {Kottmann}}, \bibinfo {author} {\bibfnamefont {Y.~W.}\
  \bibnamefont {Liu}}, \bibinfo {author} {\bibfnamefont {C.~M.~B.}\
  \bibnamefont {Monteiro}}, \bibinfo {author} {\bibfnamefont {F.}~\bibnamefont
  {Mulhauser}}, \bibinfo {author} {\bibfnamefont {T.}~\bibnamefont {Nebel}},
  \bibinfo {author} {\bibfnamefont {F.}~\bibnamefont {Nez}}, \bibinfo {author}
  {\bibfnamefont {J.~M.~F.}\ \bibnamefont {dos Santos}}, \bibinfo {author}
  {\bibfnamefont {K.}~\bibnamefont {Schuhmann}}, \bibinfo {author}
  {\bibfnamefont {D.}~\bibnamefont {Taqqu}}, \bibinfo {author} {\bibfnamefont
  {J.~F. C.~A.}\ \bibnamefont {Veloso}}, \bibinfo {author} {\bibfnamefont
  {A.}~\bibnamefont {Voss}}, \ and\ \bibinfo {author} {\bibfnamefont
  {R.}~\bibnamefont {Pohl}},\ }\href {http://dx.doi.org/10.1139/P10-113}
  {\bibfield  {journal} {\bibinfo  {journal} {Can. J. Phys.},\ }\textbf
  {\bibinfo {volume} {89}},\ \bibinfo {pages} {47} (\bibinfo {year}
  {2011})}\BibitemShut {NoStop}%
\bibitem [{\citenamefont {Nebel}\ \emph {et~al.}(2012)\citenamefont {Nebel},
  \citenamefont {Amaro}, \citenamefont {Antognini}, \citenamefont {Biraben},
  \citenamefont {Cardoso}, \citenamefont {Covita}, \citenamefont {Dax},
  \citenamefont {Fernandes}, \citenamefont {Gouvea}, \citenamefont {Graf},
  \citenamefont {H\"{a}nsch}, \citenamefont {Hildebrandt}, \citenamefont
  {Indelicato}, \citenamefont {Julien}, \citenamefont {Kirch}, \citenamefont
  {Kottmann}, \citenamefont {Liu}, \citenamefont {Monteiro}, \citenamefont
  {Nez}, \citenamefont {Santos}, \citenamefont {Schuhmann}, \citenamefont
  {Taqqu}, \citenamefont {Veloso}, \citenamefont {Voss},\ and\ \citenamefont
  {Pohl}}]{naa2012}%
  \BibitemOpen
  \bibfield  {author} {\bibinfo {author} {\bibfnamefont {T.}~\bibnamefont
  {Nebel}}, \bibinfo {author} {\bibfnamefont {F.~D.}\ \bibnamefont {Amaro}},
  \bibinfo {author} {\bibfnamefont {A.}~\bibnamefont {Antognini}}, \bibinfo
  {author} {\bibfnamefont {F.}~\bibnamefont {Biraben}}, \bibinfo {author}
  {\bibfnamefont {J.~M.~R.}\ \bibnamefont {Cardoso}}, \bibinfo {author}
  {\bibfnamefont {D.~S.}\ \bibnamefont {Covita}}, \bibinfo {author}
  {\bibfnamefont {A.}~\bibnamefont {Dax}}, \bibinfo {author} {\bibfnamefont
  {L.~M.~P.}\ \bibnamefont {Fernandes}}, \bibinfo {author} {\bibfnamefont
  {A.~L.}\ \bibnamefont {Gouvea}}, \bibinfo {author} {\bibfnamefont
  {T.}~\bibnamefont {Graf}}, \bibinfo {author} {\bibfnamefont {T.~W.}\
  \bibnamefont {H\"{a}nsch}}, \bibinfo {author} {\bibfnamefont
  {M.}~\bibnamefont {Hildebrandt}}, \bibinfo {author} {\bibfnamefont
  {P.}~\bibnamefont {Indelicato}}, \bibinfo {author} {\bibfnamefont
  {L.}~\bibnamefont {Julien}}, \bibinfo {author} {\bibfnamefont
  {K.}~\bibnamefont {Kirch}}, \bibinfo {author} {\bibfnamefont
  {F.}~\bibnamefont {Kottmann}}, \bibinfo {author} {\bibfnamefont {Y.~W.}\
  \bibnamefont {Liu}}, \bibinfo {author} {\bibfnamefont {C.~M.~B.}\
  \bibnamefont {Monteiro}}, \bibinfo {author} {\bibfnamefont {F.}~\bibnamefont
  {Nez}}, \bibinfo {author} {\bibfnamefont {J.~M. F.~d.}\ \bibnamefont
  {Santos}}, \bibinfo {author} {\bibfnamefont {K.}~\bibnamefont {Schuhmann}},
  \bibinfo {author} {\bibfnamefont {D.}~\bibnamefont {Taqqu}}, \bibinfo
  {author} {\bibfnamefont {J.~F. C.~A.}\ \bibnamefont {Veloso}}, \bibinfo
  {author} {\bibfnamefont {A.}~\bibnamefont {Voss}}, \ and\ \bibinfo {author}
  {\bibfnamefont {R.}~\bibnamefont {Pohl}},\ }\href
  {http://dx.doi.org/10.1007/s10751-012-0637-0} {\bibfield  {journal} {\bibinfo
   {journal} {Hyp. Int.},\ }\textbf {\bibinfo {volume} {212}},\ \bibinfo
  {pages} {195} (\bibinfo {year} {2012})}\BibitemShut {NoStop}%
\bibitem [{\citenamefont {Loudon}(2000)}]{rlo2000}%
  \BibitemOpen
  \bibfield  {author} {\bibinfo {author} {\bibfnamefont {R.}~\bibnamefont
  {Loudon}},\ }\href@noop {} {\emph {\bibinfo {title} {The Quantum Theory of
  Light}}}\ (\bibinfo  {publisher} {Oxford Science Publications},\ \bibinfo
  {address} {Oxford},\ \bibinfo {year} {2000})\BibitemShut {NoStop}%
\bibitem [{\citenamefont {Akhiezer}\ and\ \citenamefont
  {Berestetskii}(1965)}]{alb1965}%
  \BibitemOpen
  \bibfield  {author} {\bibinfo {author} {\bibfnamefont {A.~I.}\ \bibnamefont
  {Akhiezer}}\ and\ \bibinfo {author} {\bibfnamefont {V.~B.}\ \bibnamefont
  {Berestetskii}},\ }\href
  {http://www.amazon.com/Elements-quantum-electrodynamics-Akhiezer/dp/B0007JUZT0}
  {\emph {\bibinfo {title} {Quantum Electrodynamics}}}\ (\bibinfo  {publisher}
  {Interscience Publishers},\ \bibinfo {address} {New York},\ \bibinfo {year}
  {1965})\BibitemShut {NoStop}%
\bibitem [{\citenamefont {Safari}\ \emph {et~al.}(2014)\citenamefont {Safari},
  \citenamefont {Amaro}, \citenamefont {Santos},\ and\ \citenamefont
  {Fratini}}]{sas2014}%
  \BibitemOpen
  \bibfield  {author} {\bibinfo {author} {\bibfnamefont {L.}~\bibnamefont
  {Safari}}, \bibinfo {author} {\bibfnamefont {P.}~\bibnamefont {Amaro}},
  \bibinfo {author} {\bibfnamefont {J.~P.}\ \bibnamefont {Santos}}, \ and\
  \bibinfo {author} {\bibfnamefont {F.}~\bibnamefont {Fratini}},\ }\href
  {http://link.aps.org/doi/10.1103/PhysRevA.90.014502} {\bibfield  {journal}
  {\bibinfo  {journal} {Phys. Rev. A},\ }\textbf {\bibinfo {volume} {90}},\
  \bibinfo {pages} {014502} (\bibinfo {year} {2014})}\BibitemShut {NoStop}%
\bibitem [{\citenamefont {Pachucki}(1996)}]{pac1996}%
  \BibitemOpen
  \bibfield  {author} {\bibinfo {author} {\bibfnamefont {K.}~\bibnamefont
  {Pachucki}},\ }\href {http://link.aps.org/abstract/PRA/v53/p2092} {\bibfield
  {journal} {\bibinfo  {journal} {Phys. Rev. A},\ }\textbf {\bibinfo {volume}
  {53}},\ \bibinfo {pages} {2092} (\bibinfo {year} {1996})}\BibitemShut
  {NoStop}%
\bibitem [{Note1()}]{Note1}%
  \BibitemOpen
  \bibinfo {note} {Elliptical polarization of the incident photon can only
  decrease the QI shift and will be reported elsewhere.}\BibitemShut {Stop}%
\bibitem [{min()}]{min2015}%
  \BibitemOpen
  \href {https://root.cern.ch/root/html/TMinuit.html} {\bibinfo  {journal}
  {MINUIT - Function Minimization and Error Analysis}}\BibitemShut {NoStop}%
\bibitem [{\citenamefont {Antognini}\ \emph {et~al.}(2005)\citenamefont
  {Antognini}, \citenamefont {Amaro}, \citenamefont {Biraben}, \citenamefont
  {Cardoso}, \citenamefont {Conde}, \citenamefont {Covita}, \citenamefont
  {Dax}, \citenamefont {Dhawan}, \citenamefont {Fernandes},\ and\ \citenamefont
  {H\"{a}nsch}}]{aab2005}%
  \BibitemOpen
\bibfield  {journal} {  }\bibfield  {author} {\bibinfo {author} {\bibfnamefont
  {A.}~\bibnamefont {Antognini}}, \bibinfo {author} {\bibfnamefont {F.~D.}\
  \bibnamefont {Amaro}}, \bibinfo {author} {\bibfnamefont {F.}~\bibnamefont
  {Biraben}}, \bibinfo {author} {\bibfnamefont {J.~M.~R.}\ \bibnamefont
  {Cardoso}}, \bibinfo {author} {\bibfnamefont {C.~A.~N.}\ \bibnamefont
  {Conde}}, \bibinfo {author} {\bibfnamefont {D.~S.}\ \bibnamefont {Covita}},
  \bibinfo {author} {\bibfnamefont {A.}~\bibnamefont {Dax}}, \bibinfo {author}
  {\bibfnamefont {S.}~\bibnamefont {Dhawan}}, \bibinfo {author} {\bibfnamefont
  {L.~M.~P.}\ \bibnamefont {Fernandes}}, \ and\ \bibinfo {author}
  {\bibfnamefont {T.~W.}\ \bibnamefont {H\"{a}nsch}},\ }\href
  {http://www.sciencedirect.com/science/article/pii/S0030401805004487}
  {\bibfield  {journal} {\bibinfo  {journal} {Opt. Commun.},\ }\textbf
  {\bibinfo {volume} {253}},\ \bibinfo {pages} {362} (\bibinfo {year}
  {2005})}\BibitemShut {NoStop}%
\bibitem [{\citenamefont {Antognini}\ \emph {et~al.}(2009)\citenamefont
  {Antognini}, \citenamefont {Schuhmann}, \citenamefont {Amaro}, \citenamefont
  {Biraben}, \citenamefont {Dax}, \citenamefont {Giesen}, \citenamefont {Graf},
  \citenamefont {H\"{a}nsch}, \citenamefont {Indelicato}, \citenamefont
  {Julien}, \citenamefont {Cheng-Yang}, \citenamefont {Knowles}, \citenamefont
  {Kottmann}, \citenamefont {Le~Bigot}, \citenamefont {Yi-Wei}, \citenamefont
  {Ludhova}, \citenamefont {Moschuring}, \citenamefont {Mulhauser},
  \citenamefont {Nebel}, \citenamefont {Nez}, \citenamefont {Rabinowitz},
  \citenamefont {Schwob}, \citenamefont {Taqqu},\ and\ \citenamefont
  {Pohl}}]{asa2009}%
  \BibitemOpen
  \bibfield  {author} {\bibinfo {author} {\bibfnamefont {A.}~\bibnamefont
  {Antognini}}, \bibinfo {author} {\bibfnamefont {K.}~\bibnamefont
  {Schuhmann}}, \bibinfo {author} {\bibfnamefont {F.~D.}\ \bibnamefont
  {Amaro}}, \bibinfo {author} {\bibfnamefont {F.}~\bibnamefont {Biraben}},
  \bibinfo {author} {\bibfnamefont {A.}~\bibnamefont {Dax}}, \bibinfo {author}
  {\bibfnamefont {A.}~\bibnamefont {Giesen}}, \bibinfo {author} {\bibfnamefont
  {T.}~\bibnamefont {Graf}}, \bibinfo {author} {\bibfnamefont {T.~W.}\
  \bibnamefont {H\"{a}nsch}}, \bibinfo {author} {\bibfnamefont
  {P.}~\bibnamefont {Indelicato}}, \bibinfo {author} {\bibfnamefont
  {L.}~\bibnamefont {Julien}}, \bibinfo {author} {\bibfnamefont
  {K.}~\bibnamefont {Cheng-Yang}}, \bibinfo {author} {\bibfnamefont {P.~E.}\
  \bibnamefont {Knowles}}, \bibinfo {author} {\bibfnamefont {F.}~\bibnamefont
  {Kottmann}}, \bibinfo {author} {\bibfnamefont {E.}~\bibnamefont {Le~Bigot}},
  \bibinfo {author} {\bibfnamefont {L.}~\bibnamefont {Yi-Wei}}, \bibinfo
  {author} {\bibfnamefont {L.}~\bibnamefont {Ludhova}}, \bibinfo {author}
  {\bibfnamefont {N.}~\bibnamefont {Moschuring}}, \bibinfo {author}
  {\bibfnamefont {F.}~\bibnamefont {Mulhauser}}, \bibinfo {author}
  {\bibfnamefont {T.}~\bibnamefont {Nebel}}, \bibinfo {author} {\bibfnamefont
  {F.}~\bibnamefont {Nez}}, \bibinfo {author} {\bibfnamefont {P.}~\bibnamefont
  {Rabinowitz}}, \bibinfo {author} {\bibfnamefont {C.}~\bibnamefont {Schwob}},
  \bibinfo {author} {\bibfnamefont {D.}~\bibnamefont {Taqqu}}, \ and\ \bibinfo
  {author} {\bibfnamefont {R.}~\bibnamefont {Pohl}},\ }\href
  {http://ieeexplore.ieee.org/ielx5/3/5133711/05165103.pdf?tp=&arnumber=5165103&isnumber=5133711}
  {\bibfield  {journal} {\bibinfo  {journal} {Quantum Electronics, IEEE Journal
  of},\ }\textbf {\bibinfo {volume} {45}},\ \bibinfo {pages} {993} (\bibinfo
  {year} {2009})}\BibitemShut {NoStop}%
\bibitem [{\citenamefont {Vogelsang}\ \emph {et~al.}(2014)\citenamefont
  {Vogelsang}, \citenamefont {Diepold}, \citenamefont {Antognini},
  \citenamefont {Dax}, \citenamefont {G\"{o}tzfried}, \citenamefont
  {H\"{a}nsch}, \citenamefont {Kottmann}, \citenamefont {Krauth}, \citenamefont
  {Liu}, \citenamefont {Nebel}, \citenamefont {Nez}, \citenamefont {Schuhmann},
  \citenamefont {Taqqu},\ and\ \citenamefont {Pohl}}]{vda2014}%
  \BibitemOpen
  \bibfield  {author} {\bibinfo {author} {\bibfnamefont {J.}~\bibnamefont
  {Vogelsang}}, \bibinfo {author} {\bibfnamefont {M.}~\bibnamefont {Diepold}},
  \bibinfo {author} {\bibfnamefont {A.}~\bibnamefont {Antognini}}, \bibinfo
  {author} {\bibfnamefont {A.}~\bibnamefont {Dax}}, \bibinfo {author}
  {\bibfnamefont {J.}~\bibnamefont {G\"{o}tzfried}}, \bibinfo {author}
  {\bibfnamefont {T.~W.}\ \bibnamefont {H\"{a}nsch}}, \bibinfo {author}
  {\bibfnamefont {F.}~\bibnamefont {Kottmann}}, \bibinfo {author}
  {\bibfnamefont {J.~J.}\ \bibnamefont {Krauth}}, \bibinfo {author}
  {\bibfnamefont {Y.-W.}\ \bibnamefont {Liu}}, \bibinfo {author} {\bibfnamefont
  {T.}~\bibnamefont {Nebel}}, \bibinfo {author} {\bibfnamefont
  {F.}~\bibnamefont {Nez}}, \bibinfo {author} {\bibfnamefont {K.}~\bibnamefont
  {Schuhmann}}, \bibinfo {author} {\bibfnamefont {D.}~\bibnamefont {Taqqu}}, \
  and\ \bibinfo {author} {\bibfnamefont {R.}~\bibnamefont {Pohl}},\ }\href
  {https://www.osapublishing.org/oe/abstract.cfm?URI=oe-22-11-13050} {\bibfield
   {journal} {\bibinfo  {journal} {Optics Express},\ }\textbf {\bibinfo
  {volume} {22}},\ \bibinfo {pages} {13050} (\bibinfo {year}
  {2014})}\BibitemShut {NoStop}%
\bibitem [{\citenamefont {Ludhova}\ \emph {et~al.}(2005)\citenamefont
  {Ludhova}, \citenamefont {Amaro}, \citenamefont {Antognini}, \citenamefont
  {Biraben}, \citenamefont {Cardoso}, \citenamefont {Conde}, \citenamefont
  {Covita}, \citenamefont {Dax}, \citenamefont {Dhawan}, \citenamefont
  {Fernandes}, \citenamefont {H\"{a}nsch}, \citenamefont {Hughes},
  \citenamefont {Huot}, \citenamefont {Indelicato}, \citenamefont {Julien},
  \citenamefont {Knowles}, \citenamefont {Kottmann}, \citenamefont {Lopes},
  \citenamefont {Liu}, \citenamefont {Monteiro}, \citenamefont {Mulhauser},
  \citenamefont {Nez}, \citenamefont {Pohl}, \citenamefont {Rabinowitz},
  \citenamefont {dos Santos}, \citenamefont {Schaller}, \citenamefont {Taqqu},\
  and\ \citenamefont {Veloso}}]{laa2005}%
  \BibitemOpen
  \bibfield  {author} {\bibinfo {author} {\bibfnamefont {L.}~\bibnamefont
  {Ludhova}}, \bibinfo {author} {\bibfnamefont {F.~D.}\ \bibnamefont {Amaro}},
  \bibinfo {author} {\bibfnamefont {A.}~\bibnamefont {Antognini}}, \bibinfo
  {author} {\bibfnamefont {F.}~\bibnamefont {Biraben}}, \bibinfo {author}
  {\bibfnamefont {J.~M.~R.}\ \bibnamefont {Cardoso}}, \bibinfo {author}
  {\bibfnamefont {C.~A.~N.}\ \bibnamefont {Conde}}, \bibinfo {author}
  {\bibfnamefont {D.~S.}\ \bibnamefont {Covita}}, \bibinfo {author}
  {\bibfnamefont {A.}~\bibnamefont {Dax}}, \bibinfo {author} {\bibfnamefont
  {S.}~\bibnamefont {Dhawan}}, \bibinfo {author} {\bibfnamefont {L.~M.~P.}\
  \bibnamefont {Fernandes}}, \bibinfo {author} {\bibfnamefont {T.~W.}\
  \bibnamefont {H\"{a}nsch}}, \bibinfo {author} {\bibfnamefont {V.~W.}\
  \bibnamefont {Hughes}}, \bibinfo {author} {\bibfnamefont {O.}~\bibnamefont
  {Huot}}, \bibinfo {author} {\bibfnamefont {P.}~\bibnamefont {Indelicato}},
  \bibinfo {author} {\bibfnamefont {L.}~\bibnamefont {Julien}}, \bibinfo
  {author} {\bibfnamefont {P.~E.}\ \bibnamefont {Knowles}}, \bibinfo {author}
  {\bibfnamefont {F.}~\bibnamefont {Kottmann}}, \bibinfo {author}
  {\bibfnamefont {J.~A.~M.}\ \bibnamefont {Lopes}}, \bibinfo {author}
  {\bibfnamefont {Y.~W.}\ \bibnamefont {Liu}}, \bibinfo {author} {\bibfnamefont
  {C.~M.~B.}\ \bibnamefont {Monteiro}}, \bibinfo {author} {\bibfnamefont
  {F.}~\bibnamefont {Mulhauser}}, \bibinfo {author} {\bibfnamefont
  {F.}~\bibnamefont {Nez}}, \bibinfo {author} {\bibfnamefont {R.}~\bibnamefont
  {Pohl}}, \bibinfo {author} {\bibfnamefont {P.}~\bibnamefont {Rabinowitz}},
  \bibinfo {author} {\bibfnamefont {J.~M.~F.}\ \bibnamefont {dos Santos}},
  \bibinfo {author} {\bibfnamefont {L.~A.}\ \bibnamefont {Schaller}}, \bibinfo
  {author} {\bibfnamefont {D.}~\bibnamefont {Taqqu}}, \ and\ \bibinfo {author}
  {\bibfnamefont {J.~F. C.~A.}\ \bibnamefont {Veloso}},\ }\href
  {http://www.sciencedirect.com/science/article/pii/S0168900204024143}
  {\bibfield  {journal} {\bibinfo  {journal} {Nucl. Instrum. and Meth. Phys.
  A},\ }\textbf {\bibinfo {volume} {540}},\ \bibinfo {pages} {169} (\bibinfo
  {year} {2005})}\BibitemShut {NoStop}%
\bibitem [{\citenamefont {Fernandes}\ \emph {et~al.}(2007)\citenamefont
  {Fernandes}, \citenamefont {Amaro}, \citenamefont {Antognini}, \citenamefont
  {Cardoso}, \citenamefont {Conde}, \citenamefont {Huot}, \citenamefont
  {Knowles}, \citenamefont {Kottmann}, \citenamefont {Lopes}, \citenamefont
  {Ludhova}, \citenamefont {Monteiro}, \citenamefont {Mulhauser}, \citenamefont
  {Pohl}, \citenamefont {Santos}, \citenamefont {Schaller}, \citenamefont
  {Taqqu},\ and\ \citenamefont {Veloso}}]{faa2007}%
  \BibitemOpen
  \bibfield  {author} {\bibinfo {author} {\bibfnamefont {L.~M.~P.}\
  \bibnamefont {Fernandes}}, \bibinfo {author} {\bibfnamefont {F.~D.}\
  \bibnamefont {Amaro}}, \bibinfo {author} {\bibfnamefont {A.}~\bibnamefont
  {Antognini}}, \bibinfo {author} {\bibfnamefont {J.~M.~R.}\ \bibnamefont
  {Cardoso}}, \bibinfo {author} {\bibfnamefont {C.~A.~N.}\ \bibnamefont
  {Conde}}, \bibinfo {author} {\bibfnamefont {O.}~\bibnamefont {Huot}},
  \bibinfo {author} {\bibfnamefont {P.~E.}\ \bibnamefont {Knowles}}, \bibinfo
  {author} {\bibfnamefont {F.}~\bibnamefont {Kottmann}}, \bibinfo {author}
  {\bibfnamefont {J.~A.~M.}\ \bibnamefont {Lopes}}, \bibinfo {author}
  {\bibfnamefont {L.}~\bibnamefont {Ludhova}}, \bibinfo {author} {\bibfnamefont
  {C.~M.~B.}\ \bibnamefont {Monteiro}}, \bibinfo {author} {\bibfnamefont
  {F.}~\bibnamefont {Mulhauser}}, \bibinfo {author} {\bibfnamefont
  {R.}~\bibnamefont {Pohl}}, \bibinfo {author} {\bibfnamefont {J.~M. F.~d.}\
  \bibnamefont {Santos}}, \bibinfo {author} {\bibfnamefont {L.~A.}\
  \bibnamefont {Schaller}}, \bibinfo {author} {\bibfnamefont {D.}~\bibnamefont
  {Taqqu}}, \ and\ \bibinfo {author} {\bibfnamefont {J.~F. C.~A.}\ \bibnamefont
  {Veloso}},\ }\href {http://stacks.iop.org/1748-0221/2/i=08/a=P08005}
  {\bibfield  {journal} {\bibinfo  {journal} {Journal of Instrumentation},\
  }\textbf {\bibinfo {volume} {2}},\ \bibinfo {pages} {P08005} (\bibinfo {year}
  {2007})}\BibitemShut {NoStop}%
\bibitem [{\citenamefont {Diepold}\ \emph {et~al.}(2015)\citenamefont
  {Diepold}, \citenamefont {Fernandes}, \citenamefont {Machado}, \citenamefont
  {Amaro}, \citenamefont {Abdou-Ahmed}, \citenamefont {Amaro}, \citenamefont
  {Antognini}, \citenamefont {Biraben}, \citenamefont {Chen}, \citenamefont
  {Covita}, \citenamefont {Dax}, \citenamefont {Franke}, \citenamefont
  {Galtier}, \citenamefont {Gouvea}, \citenamefont {Götzfried}, \citenamefont
  {Graf}, \citenamefont {H\"{a}nsch}, \citenamefont {Hildebrandt},
  \citenamefont {Indelicato}, \citenamefont {Julien}, \citenamefont {Kirch},
  \citenamefont {Knecht}, \citenamefont {Kottmann}, \citenamefont {Krauth},
  \citenamefont {Liu}, \citenamefont {Monteiro}, \citenamefont {Mulhauser},
  \citenamefont {Naar}, \citenamefont {Nebel}, \citenamefont {Nez},
  \citenamefont {Santos}, \citenamefont {dos Santos}, \citenamefont
  {Schuhmann}, \citenamefont {Szabo}, \citenamefont {Taqqu}, \citenamefont
  {Veloso}, \citenamefont {Voss}, \citenamefont {Weichelt},\ and\ \citenamefont
  {Pohl}}]{dfl2015}%
  \BibitemOpen
  \bibfield  {author} {\bibinfo {author} {\bibfnamefont {M.}~\bibnamefont
  {Diepold}}, \bibinfo {author} {\bibfnamefont {L.~M.~P.}\ \bibnamefont
  {Fernandes}}, \bibinfo {author} {\bibfnamefont {J.}~\bibnamefont {Machado}},
  \bibinfo {author} {\bibfnamefont {P.}~\bibnamefont {Amaro}}, \bibinfo
  {author} {\bibfnamefont {M.}~\bibnamefont {Abdou-Ahmed}}, \bibinfo {author}
  {\bibfnamefont {F.~D.}\ \bibnamefont {Amaro}}, \bibinfo {author}
  {\bibfnamefont {A.}~\bibnamefont {Antognini}}, \bibinfo {author}
  {\bibfnamefont {F.}~\bibnamefont {Biraben}}, \bibinfo {author} {\bibfnamefont
  {T.-L.}\ \bibnamefont {Chen}}, \bibinfo {author} {\bibfnamefont {D.~S.}\
  \bibnamefont {Covita}}, \bibinfo {author} {\bibfnamefont {A.~J.}\
  \bibnamefont {Dax}}, \bibinfo {author} {\bibfnamefont {B.}~\bibnamefont
  {Franke}}, \bibinfo {author} {\bibfnamefont {S.}~\bibnamefont {Galtier}},
  \bibinfo {author} {\bibfnamefont {A.~L.}\ \bibnamefont {Gouvea}}, \bibinfo
  {author} {\bibfnamefont {J.}~\bibnamefont {Götzfried}}, \bibinfo {author}
  {\bibfnamefont {T.}~\bibnamefont {Graf}}, \bibinfo {author} {\bibfnamefont
  {T.~W.}\ \bibnamefont {H\"{a}nsch}}, \bibinfo {author} {\bibfnamefont
  {M.}~\bibnamefont {Hildebrandt}}, \bibinfo {author} {\bibfnamefont
  {P.}~\bibnamefont {Indelicato}}, \bibinfo {author} {\bibfnamefont
  {L.}~\bibnamefont {Julien}}, \bibinfo {author} {\bibfnamefont
  {K.}~\bibnamefont {Kirch}}, \bibinfo {author} {\bibfnamefont
  {A.}~\bibnamefont {Knecht}}, \bibinfo {author} {\bibfnamefont
  {F.}~\bibnamefont {Kottmann}}, \bibinfo {author} {\bibfnamefont {J.~J.}\
  \bibnamefont {Krauth}}, \bibinfo {author} {\bibfnamefont {Y.-W.}\
  \bibnamefont {Liu}}, \bibinfo {author} {\bibfnamefont {C.~M.~B.}\
  \bibnamefont {Monteiro}}, \bibinfo {author} {\bibfnamefont {F.}~\bibnamefont
  {Mulhauser}}, \bibinfo {author} {\bibfnamefont {B.}~\bibnamefont {Naar}},
  \bibinfo {author} {\bibfnamefont {T.}~\bibnamefont {Nebel}}, \bibinfo
  {author} {\bibfnamefont {F.}~\bibnamefont {Nez}}, \bibinfo {author}
  {\bibfnamefont {J.~P.}\ \bibnamefont {Santos}}, \bibinfo {author}
  {\bibfnamefont {J.~M.~F.}\ \bibnamefont {dos Santos}}, \bibinfo {author}
  {\bibfnamefont {K.}~\bibnamefont {Schuhmann}}, \bibinfo {author}
  {\bibfnamefont {C.~I.}\ \bibnamefont {Szabo}}, \bibinfo {author}
  {\bibfnamefont {D.}~\bibnamefont {Taqqu}}, \bibinfo {author} {\bibfnamefont
  {J.~F. C.~A.}\ \bibnamefont {Veloso}}, \bibinfo {author} {\bibfnamefont
  {A.}~\bibnamefont {Voss}}, \bibinfo {author} {\bibfnamefont {B.}~\bibnamefont
  {Weichelt}}, \ and\ \bibinfo {author} {\bibfnamefont {R.}~\bibnamefont
  {Pohl}},\ }\href
  {http://scitation.aip.org/content/aip/journal/rsi/86/5/10.1063/1.4921195
  http://scitation.aip.org/docserver/fulltext/aip/journal/rsi/86/5/1.4921195.pdf?expires=1432162595&id=id&accname=guest&checksum=B4D00D640D8547979336A725FED0E3BB}
  {\bibfield  {journal} {\bibinfo  {journal} {Rev. Sci. Instrum.},\ }\textbf
  {\bibinfo {volume} {86}},\ \bibinfo {pages} {053102} (\bibinfo {year}
  {2015})}\BibitemShut {NoStop}%
\bibitem [{\citenamefont {Rose}(1957)}]{ros1957}%
  \BibitemOpen
  \bibfield  {author} {\bibinfo {author} {\bibfnamefont {M.~E.}\ \bibnamefont
  {Rose}},\ }\href@noop {} {\emph {\bibinfo {title} {Elementary Theory of
  Angular Momentum}}}\ (\bibinfo  {publisher} {John Wiley},\ \bibinfo {address}
  {New York},\ \bibinfo {year} {1957})\BibitemShut {NoStop}%
\end{thebibliography}%

\end{document}